\documentclass[prl, twocolumn, superscriptaddress, amsmath, amssymb, floatfix, longbibliography, nofootinbib, notitlepage]{revtex4-2}
\usepackage{amsbsy}
\usepackage{amsfonts}
\usepackage{amsmath}
\usepackage{amstext}
\usepackage{amsthm}
\usepackage{amssymb}
\usepackage{graphicx}
\usepackage{epsfig}
\usepackage{enumerate}
\usepackage{epstopdf}
\usepackage{bm}
\usepackage{bbm}
\usepackage{braket}
\usepackage{color}
\usepackage{subcaption}
\definecolor{prlblue}{rgb}{0.18,0.19,0.57}
\usepackage{multirow}
\usepackage[colorlinks]{hyperref}
\usepackage{cleveref}
\usepackage{ifthen}
\usepackage{tikz}
\usepackage{graphicx} % Required for inserting images
\usetikzlibrary{arrows.meta,calc,positioning,decorations.pathreplacing}
\tikzset{
  tensor/.style={draw, rounded corners=2pt, minimum size=12mm, thick, fill=gray!10},
  leg/.style={thick},
  op/.style={draw, circle, inner sep=1.2pt, thick, fill=white},
  every node/.style={font=\small}
}
 
\usetikzlibrary{positioning,calc}
\usepackage{ amsfonts, mathtools, physics, float,bm,url,tabularx,bbm,slashed,xcolor}
\usepackage [english]{babel}
\usepackage [autostyle, english = american]{csquotes}
\usepackage{breqn}
\usepackage{titlesec}
\usepackage{enumerate}
\usepackage{epstopdf}
\usepackage{booktabs}
\MakeOuterQuote{"}
\graphicspath{ {./images/} }

\newcommand\mathnorm[1]{\left\lVert#1\right\rVert}
\newcommand{\Z}{\mathbb{Z}}

\newcommand{\ie}{\begin{equation}\begin{aligned}}
\newcommand{\fe}{\end{aligned}\end{equation}}
\newcommand{\bbra}[1]{\left( #1 \right\lvert}
\newcommand{\bket}[1]{\left\lvert #1 \right)}

\newcommand{\newsec}[1]{\textcolor{blue}{{\textit{#1}.-}}}

\newtheorem{lemma}{Lemma}
\newtheorem*{claim*}{Claim}
\theoremstyle{remark}
\newtheorem*{remark*}{Remark}

\begin{document}
\title{Matrix Product States for Modulated Symmetries: SPT, LSM, and Beyond}
\author{Amogh Anakru}
\thanks{These authors contributed equally to this work.}
%\email{apa6106@psu.edu}
\affiliation{Department of Physics$,$ The Pennsylvania State University$,$ University Park$,$ Pennsylvania$,$ 16802$,$ USA}

\author{Linhao Li}
\thanks{These authors contributed equally to this work.}
\email{linhaoli601@gmail.com}
\affiliation{Department of Physics$,$ The Pennsylvania State University$,$ University Park$,$ Pennsylvania$,$ 16802$,$ USA}

\author{Sarvesh Srinivasan}
%\email{sbs5627@psu.edu}
\affiliation{Department of Physics$,$ The Pennsylvania State University$,$ University Park$,$ Pennsylvania$,$ 16802$,$ USA}
\author{Zhen Bi}
\email{zjb5184@psu.edu}
\affiliation{Department of Physics$,$ The Pennsylvania State University$,$ University Park$,$ Pennsylvania$,$ 16802$,$ USA}
\affiliation{Center for Theory of Emergent Quantum Matter$,$ Institute for Computational and Data Sciences$,$ The Pennsylvania State University$,$ University Park$,$ Pennsylvania$,$ 16802$,$ USA}
\begin{abstract}
Matrix product states (MPS) provide a powerful framework for characterizing one-dimensional symmetry-protected topological (SPT) phases of matter and for formulating Lieb–Schultz–Mattis (LSM)-type constraints. Here we generalize the MPS formalism to translationally invariant systems with general modulated symmetries. We show that the standard symmetry “push-through” condition for conventional global symmetry must be revised to account for symmetry modulation, and we derive the appropriate generalized condition. Using this generalized push-through structure, we classify one-dimensional SPT phases with modulated symmetries and formulate LSM-type constraints within the same MPS-based framework.
\end{abstract}

\maketitle

\newsec{Introduction}
The interplay between symmetries and topology has long provided a powerful lens for discovering and classifying quantum phases of matter. In one-dimensional systems, matrix product states (MPS) have emerged as the unifying framework for understanding this interplay \cite{perez2006matrix, pollmann2010entanglement}. By efficiently representing the local entanglement structure of quantum states in 1D, MPS provides a direct bridge between microscopic lattice symmetries and macroscopic topological phenomena. Two great successes of this framework are the classification of symmetry-protected topological (SPT) phases and the formulation of generalized Lieb-Schultz-Mattis (LSM) theorems. For SPT phases---states that are gapped and non-trivial only in the presence of specific symmetries---the MPS formalism explicitly reveals how physical symmetries ``push through'' the bulk to manifest as localized projective representations at the boundaries \cite{chen2011classification, schuch2011classifying,PhysRevB.85.075125,PhysRevB.87.155114}. Conversely, the MPS framework natively captures LSM anomalies, dictating when microscopic symmetry conditions strictly forbid a unique, featureless, gapped ground state, instead forcing the system to be either gapless, symmetry-broken, or topologically ordered \cite{lieb1961two, oshikawa2000commensurability, hastings2004lieb, parameswaran2013topological,PhysRevB.93.104425,watanabe2015filling,ogata2021general,PhysRevLett.126.217201,10.21468/SciPostPhys.13.3.066,10.21468/SciPostPhys.15.2.051, PhysRevB.110.045118,PhysRevB.106.224420,PhysRevLett.133.136705}.

Recently, a new frontier has opened with the study of systems governed by \textit{modulated symmetries}---conservation laws whose unitary action varies in space. Such symmetries naturally arise in diverse experimental contexts, ranging from center-of-mass conservation in the lowest Landau level of the quantum Hall effect \cite{seidel2005incompressible} to strongly tilted optical lattices in cold atom setups \cite{guardado2020subdiffusion, scherg2021observing}. More broadly, modulated symmetries are the building blocks of fractonic matter \cite{chamon2005quantum, haah2011local, vijay2016fracton, nandkishore2019fractons, pretko2020fracton, shirley2018foliated}, where exotic mobility restrictions emerge directly from local symmetry constraints \cite{radicevic2018fractons, gromov2019towards, you2018subsystem, seiberg2020fractons,sala2022dynamics_modulated}. While recent theoretical efforts have made much progress exploring the topological phases associated with these symmetries \cite{lake_Dipolar_SPT,ye2022lieb, oh2022rank, pace2022position,delfino2023effective,watanabe_exponential_toric_code,suzuki2024lieb, ho_tat_lam_dipolar_SPT_classification, fuji2024topological,yan2024generalized,10.21468/SciPostPhys.17.4.104,ho_tat_lam_dipolar_SPT_classification,kim_you_han_noninvertible_holography_mspt,yao2025lattice,Ebisu:2025mtb,delfino_gauging_exponential,Li:2026aja}, a unified MPS study of SPT classifications and LSM constraints for \textit{arbitrary} spatial modulations has remained unavailable.

In this work, we generalize the MPS framework to translationally invariant systems with arbitrary discrete modulated symmetries. By deriving a generalized “push-through” condition, we determine how spatially modulated symmetry operations act on the virtual legs of injective MPS states. This framework enables a classification of 1D SPT phases protected by modulated symmetries, and it also yields generalized LSM and SPT–LSM-type constraints \cite{yang2018dyonic,jiang2021generalized,LU2024169806}. Finally, we propose lattice models that exhibit the corresponding LSM constraints.

\newsec{Generalized symmetry push-through condition} The ground state of a one-dimensional gapped, translation-invariant Hamiltonian with periodic boundary conditions can, in general, be represented by an injective MPS,
\begin{equation}\label{eq:MPS_DEF_main}
\ket{\psi} = \sum_{s_1\ldots s_L}\Tr(A^{s_1}\dots A^{s_L})\ket{s_1\ldots s_L},
\end{equation}
where $A^{s}$ are finite-dimensional square matrices, taken to be identical on every site.
Mathematically, injectivity demands that the transfer matrix $\mathbb{E} = \sum_s A^{s} \otimes (A^s)^*$ has a unique dominant eigenvalue separated by a spectral gap. This condition physically ensures that all correlation functions decay exponentially and the state is the non-degenerate ground state of a gapped local Hamiltonian.

To classify 1D SPT phases protected by global symmetries, it is sufficient to analyze how the symmetry acts on a single tensor. For injective MPS, the unitary transformation on the physical degrees of freedom can be ``pushed through'' the tensor and represented as unitaries acting on the virtual bonds. For unmodulated symmetries, it is sufficient to assume symmetry actions on two virtual legs are conjugate to each other \cite{cirac_pushthrough, pedagogical_pushthrough}. Importantly, while the physical degrees of freedom carry a linear representation of the symmetry, the virtual degrees of freedom may transform projectively under the symmetry, characterized by a 2-cocycle $\omega\in H^2(G,U(1))$. This cocycle is invariant under symmetric finite-depth local unitary circuits and thus classifies 1D SPT phases \cite{chen2011classification, schuch2011classifying,PhysRevB.85.075125,PhysRevB.87.155114}.\footnote{More precisely, this classification captures the strong SPTs protected by the global symmetry.} The MPS framework can also yield LSM-type constraints: if the physical degrees of freedom carry projective representations that are incompatible with symmetry push-through conditions for an injective MPS, then a symmetric short-range-entangled gapped phase is forbidden.

Here we address the classification of 1D SPT phases and LSM constraints for \emph{modulated} symmetries with a \emph{discrete} on-site group $G$. To define a modulated symmetry, we consider a translationally invariant chain of length $L$ with periodic boundary conditions. A modulated symmetry action is a unitary of the form $\mathcal{U}_g=\bigotimes_{j=1}^{L} U_{g,j}$, where $g\in G$ and the site dependence can be encoded by a modulation function $f(j)$ via $U_{g,j}=U_g^{f(j)}$. Generally, the system may admit a set of allowed modulation functions. Translation invariance imposes that this set must be closed under spatial shifts: translating a symmetry operator must produce another allowed symmetry operator. For example, the dipole modulation $f_d(j)=j$ is not shift-closed by itself; consistency requires also the global charge symmetry $f_c(j)=1$\cite{gromov2019}. Such a closed set of modulated functions defines a consistent modulated symmetry. Equivalently, a modulated symmetry may be specified by how translation acts on the on-site symmetry group $G$~\cite{bulmash2025defect, pace_bulmash_2026_LSM_Modulated}. Conjugation by the translation generator $\mathcal{T}$ defines an automorphism $\mathcal{T}(g)$ of $G$. On a periodic chain of length $L$, consistency requires $\mathcal{T}^L(g)=g$. This structure is also known as an \emph{$L$-cycle symmetry}~\cite{Lcycle2019} in the MPS literature. 

Our goal is to formulate SPT classifications and LSM-type constraints for modulated symmetries within the MPS framework. We therefore consider translationally invariant injective MPS satisfying $\mathcal{U}_g\ket{\psi}=e^{i\theta_g}\ket{\psi}$. To obtain a well-defined thermodynamic limit, we further assume that for every positive integer $k$, the MPS on a chain of length $kL$, formed by concatenating $k$ identical $L$-site unit cells, is also an eigenstate of the $k$-fold periodic extension of the symmetry $ \bigotimes^k\mathcal{U}_g$. This ensures that the modulated symmetry structure extends consistently to arbitrary large system size.

The key question is how a modulated symmetry acting on the physical indices pushes through to the virtual bonds. Assuming injectivity and a well-defined thermodynamic limit, we show that the modulated unitary action on the physical leg necessarily pushes through to the virtual legs (see Supplemental Material for details). The essential difference from the unmodulated case is that, because the physical symmetry depends on the site index, the two virtual legs of an MPS tensor need not transform identically. Instead, the symmetry is absorbed into \emph{site-dependent} unitaries $v_{j-1}$ and $v_j$ acting on the virtual legs, which encode bond ``charges'' that may differ on the left and right bonds. Concretely, one has

\begin{equation}
\label{eq:general_push_through_main}
    U_j \cdot A \doteq \, v_{j-1}^\dagger\, A\, v_j,
\end{equation}
which should be understood as the tensor equation illustrated in Fig.~\ref{fig:pushthrough-result_main}; here ``$\doteq$'' denotes equality up to a phase, and the group label is suppressed when no confusion can arise. The same push-through condition has also appeared in the MPS and MBQC literature~\cite{stephen2019subsystem,sahay2025classifying,Stephen_2025,measurement_and_feedback_girvin,sarang_measurement_feedback}; our derivation follows a different strategy and also applies to certain classes of MPS with open boundary conditions. Translation invariance further imposes a constraint on the bond unitaries, namely 
\begin{equation}\label{eq:bond_constraint_main}
    v_j(g)\doteq\,v_{j-1}(\mathcal{T}(g)).
\end{equation}

\begin{figure}[h]
\centering
\begin{tikzpicture}[
x=0.85cm,y=0.85cm,
tensor/.style={draw, rounded corners=1.5pt, minimum size=7.5mm, thick, fill=gray!10},
leg/.style={thick},
op/.style={draw, circle, inner sep=0.6pt, thick, fill=white},
every node/.style={font=\scriptsize},
]
\coordinate (M) at (3,0);

\node[tensor] (B1) at ($(M)+(0,0)$) {$A$};
\draw[leg] (B1.north) -- +(0,0.4) node[op,above] {$\ \ U_j\ \ $} -- +(0.,1.6);
\draw[leg] (B1.west)  -- +(-0.85,0);
\draw[leg] (B1.east)  -- +(0.85,0);

\node[font=\small] (eq2) at ($(M)+(1.9,0)$) {$\doteq$};

\node[tensor] (B2) at ($(M)+(4.9,0)$) {$A$};
\draw[leg] (B2.north) -- +(0,1.4);
\draw[leg] (B2.west)  -- +(-0.4,0) node[op,left] {$v_{j-1}^\dag$}  -- +(-2,0);
\draw[leg] (B2.east)  -- +(0.4,0)  node[op,right] {$\ \ v_j\ \ $} -- +(+2,0);
\end{tikzpicture}
\caption{Generalized push-through rule for modulated unitary symmetries. }
\label{fig:pushthrough-result_main}
\end{figure}

We now discuss the symmetry representations carried by the physical and virtual degrees of freedom. For SPT phases, we assume that the physical Hilbert space transforms linearly on each site, while the virtual bonds may carry projective representations characterized by a 2-cocycle $\omega_j$ defined through $v_j(g) v_j(h)=\omega_j(g,h)\,v_j(gh)$. Eq.~\eqref{eq:bond_constraint_main} then implies that, up to coboundaries, the cocycle is site-independent and satisfies
\begin{equation}\label{eq:Grand_Pullback_Cocycle_Constraint_main}
\omega(g,h)=\omega(\mathcal{T}(g),\mathcal{T}(h)).
\end{equation}
We note that, by carefully tracking the phases in Eqs.~\eqref{eq:general_push_through_main} and \eqref{eq:bond_constraint_main}, one can also identify the so-called weak SPT phases, as illustrated in Supplemental Material.

For LSM-type constraints, we allow the physical degrees of freedom to transform projectively, with cocycle $\nu_j\in H^2(G,U(1))$ defined by $U_{g,j}U_{h,j}=\nu_j(g,h)\,U_{gh,j}$. Translation invariance requires $\nu_j(g,h)=\nu_{j-1}(\mathcal{T}(g),\mathcal{T}(h))$. If a translation invariant injective MPS exists with these symmetry representations, then the physical and virtual projective data must satisfy
\begin{equation}\label{eq:Grand_SPT_LSM intro}
\begin{split}
\omega_j(g,h) &= \nu_j(g,h)\,\omega_{j-1}(g,h),\\
\omega_j(g,h) &= \omega_{j-1}(\mathcal{T}(g),\mathcal{T}(h)),
\end{split}
\end{equation}
up to coboundaries. Conversely, if these conditions fail, then no injective MPS is compatible with the symmetry representations, yielding an LSM-type constraint.

These equations serve as the engine for the remainder of this work. It allows us to classify SPTs, and identify LSM constraints with modulated symmetries. 

\newsec{SPT classification with modulated symmetries} We now present two examples -- (i) an exponential symmetry and (ii) two exponential symmetries with different exponents -- to illustrate how to classify SPTs using Eq.~\eqref{eq:Grand_Pullback_Cocycle_Constraint_main}.

We first consider an exponential symmetry \cite{hu2024quantumbreakdownmodellattice,hu2025quantumbreakdowncondensatedisorderfree}, generated by
$\mathcal{U}_g=\bigotimes_{j=1}^{L} U_g^{\,b^{j-1}}$.
Assuming $U_g$ has order $N$, periodic boundary condition requires $U_g^{\,b^L-1}=\mathbbm{1}$, which in turn implies that $b$ must be coprime to $N$.\footnote{If $b$ is not coprime to $N$, then $U_g^{b^L}=U_g$ cannot hold, since $U_g^{b^L}$ would necessarily have order smaller than $N$. By contrast, if $b$ is coprime to $N$, one can always find allowed system sizes $L$, for instance multiples of Euler's totient function $\varphi(N)$.} Applying the generalized push-through condition, we obtain
\begin{equation}\label{eq:exponential-push-through-main}
    U_g \cdot A \doteq\, v_g^\dagger\, A\, v_g^{\,b}.
\end{equation}

The SPT classification problem is to determine which projective representations can be carried by the virtual bond unitary $v_g$. Since $\mathcal{T}\mathcal{U}_g\mathcal{T}^\dagger=\mathcal{U}_g^b$, or equivalently $\mathcal{T}(g)=g^b$, Eq.~\eqref{eq:Grand_Pullback_Cocycle_Constraint_main} requires the virtual cocycle to satisfy
\begin{equation}\label{eq: exp SPT constra}
    \omega(g,h)=\omega(g^b,h^b)=\omega(g,h)^{b^2},
\end{equation}
up to coboundaries. This condition selects the projective classes that can be consistently realized on the virtual bonds. The result is in agreement with results in Ref.~\cite{pace_bulmash_2026_LSM_Modulated} derived using a cellular chain complex formalism.

Take $G=\mathbb{Z}_N\times\mathbb{Z}_N$ as an example. Since $H^2(G,U(1))=\mathbb{Z}_N$, projective representations are labeled by $k\in\mathbb{Z}_N$, which may be extracted from the coboundary-invariant phase $\omega(g,h)\omega(h,g)^{-1}=\exp(2\pi i k/N)$, with $g,h$ the generators of the two $\mathbb{Z}_N$ factors. Eq. ~\eqref{eq: exp SPT constra} restricts $k$ to multiples of $N/\gcd(b^2-1,N)$, where $\gcd$ represents the greatest common divisor. Therefore, the SPT phases protected by $\mathbb{Z}_N\times\mathbb{Z}_N$ exponential symmetry are classified by $\mathbb{Z}_{\gcd(b^2-1,N)}$.

As a second example, we consider SPT phases protected by two exponential symmetries with different exponents. 
Concretely, we take 
\begin{equation}\label{eq:exp_charge_symmetry}
\mathcal{U}_{g_a} = \bigotimes_{j=1}^L U_{g_a}^{a^{j-1}},\qquad
\mathcal{U}_{g_b} = \bigotimes_{j=1}^L U_{g_b}^{b^{j-1}},
\end{equation}
with $g_{a/b}\in G_{a/b}$. The unitary operators associated with these two symmetries on the virtual bond are denoted by $v_a(g_a)$ and $v_{b}(g_b)$. Their projective data contain three sectors: $a-a$ cocycles $\omega_a(g_a,h_a)$, $b-b$ cocycles $\omega_b(g_b,h_b)$, and mixed $a-b$ cocycles $\phi(g_a,g_b)$, here $g_{a/b},h_{a/b} \in G_{a/b}$. Applying Eq.~\eqref{eq:Grand_Pullback_Cocycle_Constraint_main}, we find that these cocycles must satisfy
\begin{equation}
\label{eq:c-e constraint_main}
    \omega_a^{\,a^2-1}=1,\qquad \phi^{\,ab-1}=1,\qquad \omega_b^{\,b^2-1}=1
\end{equation}
up to coboundary transformations.

As a simple example, let $G_a=G_b=\mathbb{Z}_N$, and $a=1$. This corresponds to one global symmetry $G_a$ and one exponential symmetry $G_b$. In this case, the $a-a$ and $b-b$ cocycles are trivial, so the only nontrivial projective data arise from the mixed cocycle $\phi(g_a,g_b)$ between the global charge and exponential symmetries. This mixed cocycle class is characterized by the coboundary-invariant phase
$\phi(g_a,g_b)\phi(g_b,g_a)^{-1}
=\exp\!\left(\frac{2\pi i k\, g_a g_b}{N}\right)$,
where $k,g_a,g_b\in\mathbb{Z}_N$. Eq.~\eqref{eq:c-e constraint_main} then restricts the allowed cocycles to those for which $k$ is a multiple of $N/\gcd(b-1,N)$. It follows that the SPT phases protected by charge and exponential symmetries are classified by $\mathbb{Z}_{\gcd(b-1,N)}$. In End matter, we construct explicit MPS representations realizing every SPT phase in this classification.

We note an interesting subtlety when the two exponential symmetries share the same on-site group, for example when both are embedded in a single $\mathbb{Z}_N$. In this case, the two symmetries need not be independent. For example, when $a=1$, the symmetry group is not truly $\mathbb{Z}_N\times\mathbb{Z}_N$, but is reduced to $\mathbb{Z}_N\times \mathbb{Z}_{N/\text{gcd}(b-1,N)}$, since certain exponential elements collapse to a uniform charge symmetry. As a result, the SPT classification needs to be refined. 

The formalism also applies directly to polynomial symmetries, such as dipole and multipole symmetries, reproducing known classifications (see Supplemental Material).

\newsec{Open boundaries and finite-size edge degeneracies}
All of the above concerns a periodic chain, on which the modulated symmetry is well defined only for special lengths: closure under periodic boundary conditions demands the $L$-cycle condition $T^{L}=\mathrm{id}$. An open chain removes this restriction---one can write a Hamiltonian invariant under the modulated symmetry for \emph{any} $L$. Being an SPT, the system then hosts edge modes. For a generic 1D SPT these edge degeneracies are exact only in the thermodynamic limit and split exponentially, $\sim e^{-\gamma L}$, on a finite chain. As a distinctive feature of modulated SPTs, we observe that this splitting can instead be symmetry-forbidden: the edge degeneracy can be exact at finite size for certain $L$.

\begin{figure}
    \includegraphics[width=0.9\linewidth]{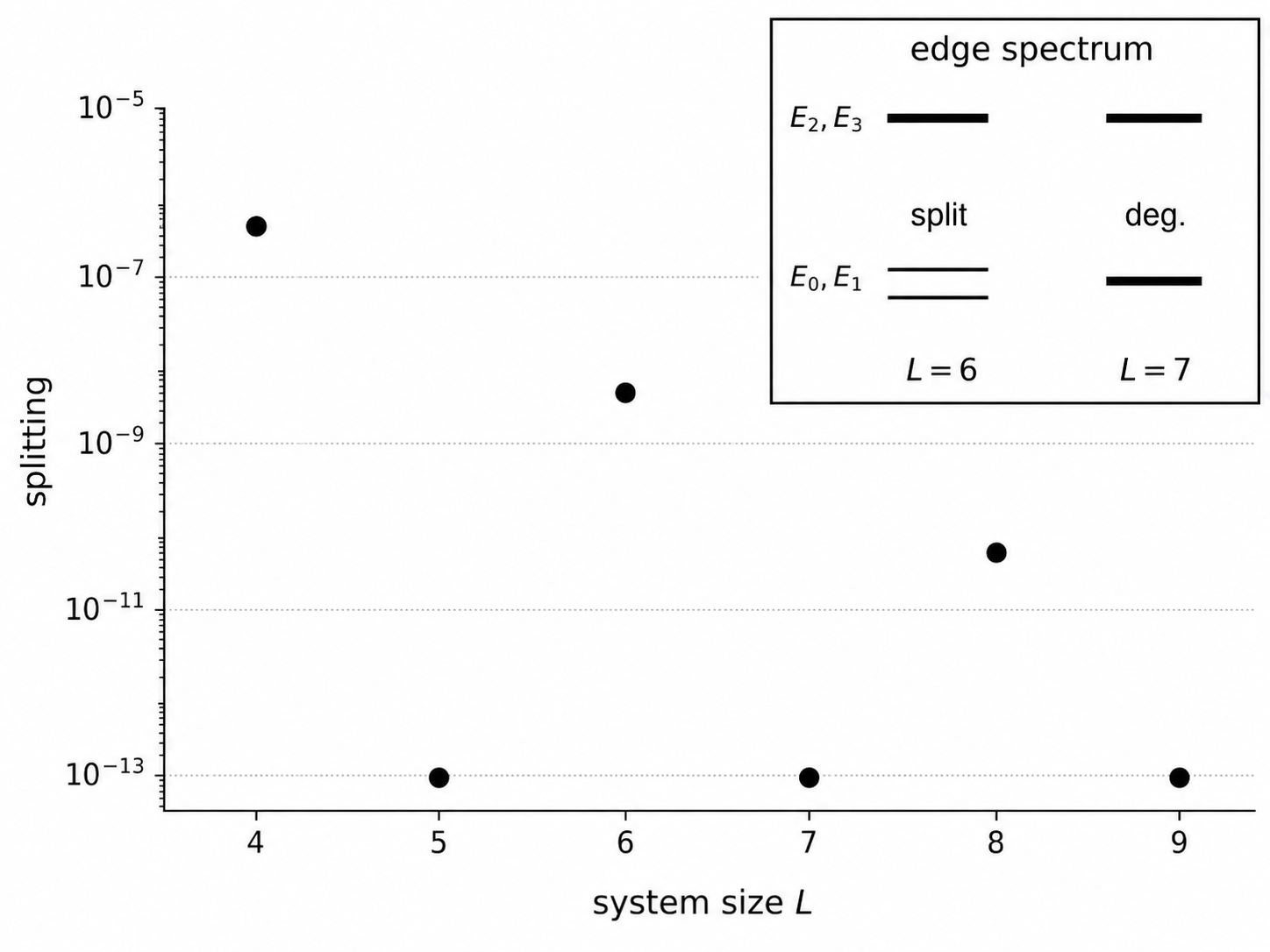}
    \caption{Finite-size splitting of the edge-state doublet for the Hamiltonian in Eq.~\ref{Eq:Exp_Chain_Numerical_Z8}, with $J_1=J_2=0.4$, and $\eta=0.5$. Inset: schematic low-energy edge spectrum for representative even and odd sizes. At $L=7$, the edge spectrum consists of two degenerate doublets within numerical precision, whereas at $L=6$ one doublet is split. Odd-$L$ splittings that fall at the numerical precision floor are shown at $10^{-13}$ for clarity.
}
    \label{fig:main_splitting_fig}
\end{figure}

To exhibit the effect concretely, we study a spin chain with one spin-1/2 and one spin-3/2 per unit cell, realizing a $D_{4N}=\mathbb{Z}_{2N}\rtimes\mathbb{Z}_{2}$ SPT with the $\mathbb{Z}_{2N}$ exponentially modulated with base $b=3$. For $2N=8$ the Hamiltonian reads (a general model is presented in End Matter)
\begin{equation}\label{Eq:Exp_Chain_Numerical_Z8}
\begin{split}
  H = &-\sum_{j}\!((\hat{L}_{j}^{+})^{3}\hat{s}^{-}_{j+1}+\text{h.c.})
       -J_{1}\sum_{j}\!(\hat{L}^{-}_{j}\hat{s}^{+}_{j}+\text{h.c.})\\
      &-J_{2}\sum_{j}\!\left(\hat{s}^{z}_{j}+\hat{L}^{z}_{j}\right)^{2}
       -\eta\sum_{j}\!(\hat{s}^{-}_{j}\hat{L}^{+}_{j+2}+\text{h.c.}).
\end{split}
\end{equation}
The first three terms preserve an exponential U(1) symmetry generated by $\mathcal{U}_1(\theta)=\exp\{i\theta\sum_j 3^j(\hat{s}^z_j+\hat{L}^z_j)\}$ and a $\mathbb{Z}_2$ given by spin-flip labeled by $\mathcal{U}_2$. The $\eta$-term breaks U(1) to the $\mathbb{Z}_{8}$ subgroup generated by $\mathcal{U}_1(\pi/4)$. Exact diagonalization (Fig.~\ref{fig:main_splitting_fig}) shows the degeneracy behaving unlike a generic SPT: rather than splitting as $\sim e^{-\gamma L}$ at all finite $L$, it remains \emph{exactly} degenerate at odd $L$.

To understand the degeneracy, we inspect the representation the two edges carry. On the left edge it is generated by $v_{1}=e^{i\theta\sigma_{z}/2}$ ($\theta=\pi/N$) and $v_{2}=\sigma_{x}$; modulation replaces the right-edge $\mathbb{Z}_{2N}$ generator by $v_{1}^{\,b^{L}}$, with $v_{2}$ unchanged. Writing the edge states $|\psi_{\alpha\beta}\rangle$ with $\alpha,\beta=\pm1$ the $\sigma_{z}$ eigenvalues of the left and right edge, the operator $\mathcal{U}_{2}$ pairs $|\psi_{\alpha\beta}\rangle\!\leftrightarrow\!|\psi_{-\alpha,-\beta}\rangle$, and on this pair
\begin{equation}
  \mathcal{U}_1(\pi/N)\mapsto\begin{pmatrix}q&\\&q^{-1}\end{pmatrix},\qquad
  \mathcal{U}_{2}\mapsto\sigma_{x},
  \label{eq:dihedral-doublet}
\end{equation}
with $q=e^{\,i(b^{L}\beta-\alpha)\theta/2}$. This two-dimensional representation of $D_{4N}$ is irreducible---and the doublet exactly degenerate---unless $q=q^{-1}$, i.e. unless $b^{L}\beta-\alpha\equiv 0\pmod{2N}$, in which case it reduces to two one-dimensional representations and may split. The edge degeneracy is thus pinned whenever this condition fails; for $2N=8$ and $b=3$ it
holds only at even $L$, matching the splitting seen in Fig.~\ref{fig:main_splitting_fig}.

\newsec{LSM theorems with modulated symmetries}
To derive LSM-type theorems within the MPS framework, we ask whether a given assignment of on-site symmetry representations is compatible with the injective MPS conditions in Eq.~\eqref{eq:Grand_SPT_LSM intro}. We illustrate this procedure with two examples: (i) two exponential symmetries and (ii) a non-Abelian symmetry group with an exponential structure. 

In the first example, we consider a $\mathbb{Z}_N$ qudit chain. We  introduce two exponential symmetries,
\begin{equation}
    \mathcal{U}_x=\bigotimes_{j=1}^L X_j^{a^{j-1}}, \qquad
    \mathcal{U}_z=\bigotimes_{j=1}^L Z_j^{\,n b^{j-1}},
\end{equation}
where $a$ and $b$ are coprime to $N$. The corresponding local symmetry actions at site $j$ furnish a projective representation of $\mathbb{Z}_N\times\mathbb{Z}_N$, characterized by the coboundary-invariant phase
$\nu_j(z,x)\nu_j(x,z)^{-1}
=\exp\!\left[\frac{2\pi i n (ab)^{j-1}}{N}\right]$,
since the on-site symmetry operators commute only up to this phase. Such a site-dependent projective representation pattern is required to be consistent with the exponential symmetry and translation invariance.

Suppose there exists an injective MPS preserving these symmetries. The generalized push-through condition then implies
\begin{equation}
\begin{split}
   X^{a^{j-1}}\cdot A&\doteq (v_x^{\dagger})^{a^{j-1}} A v_x^{a^j}, \\
 Z^{n b^{j-1}}\cdot A&\doteq (v_z^{\dagger})^{b^{j-1}} A v_z^{b^j},
\end{split}
\end{equation}

where $v_x$ and $v_z$ act on the virtual indices. 
The corresponding virtual projective class is characterized by the coboundary-invariant phase $\omega_j(z,x)\omega_j(x,z)^{-1}=v_z^{b^j}v_x^{a^j}v_z^{-b^j}v_x^{-a^j}$. If we assume the starting point is $\omega_0(z,x)\omega_0(x,z)^{-1}=v_zv_xv_z^{-1}v_x^{-1}=e^{2\pi i k/N}$ with $k\in\mathbb{Z}_N$, one obtains for site $j$: $\omega_j(z,x)\omega_j(x,z)^{-1}=e^{2\pi i k (ab)^j/N}$. Eq.~\eqref{eq:Grand_SPT_LSM intro} requires $\omega_j(z,x)\omega_j(x,z)^{-1}=\omega_{j-1}(z,x)\omega_{j-1}(x,z)^{-1}\nu_j(z,x)\nu_j(x,z)^{-1}$, which gives $k[(ab)^j-(ab)^{j-1}]=n(ab)^{j-1}\pmod N$. Since $a$ and $b$ are coprime to $N$, this reduces to
\begin{equation}
    k(ab-1)=n \pmod N.
\end{equation}
This is the necessary condition for the existence of a symmetric injective MPS, in agreement with the group-cohomology result of Ref.~\cite{pace_bulmash_2026_LSM_Modulated}. Let $c=\gcd(ab-1,N)$. Then solutions exist iff $c\mid n$, in which case
\begin{equation}\label{eq: solution}
k=\frac{n}{c}\Bigl(\frac{ab-1}{c}\Bigr)^{-1}_{N/c}+\tau\,\frac{N}{c},
\qquad \tau\in\mathbb{Z}_c,
\end{equation}
where $(\cdots)^{-1}_{N/c}$ denotes the multiplicative inverse modulo $N/c$. For $n=0$, this reduces to the SPT classification discussed above.

If $c\nmid n$, no solution exists, and an LSM constraint follows. Even when $c\mid n$, an obstruction to a trivial phase may remain: if every allowed solution has $k\neq 0 \pmod N$, the system is forced into a nontrivial SPT phase, yielding an SPT-LSM theorem \cite{yang2018dyonic,jiang2021generalized,LU2024169806}. In particular, when $c=1$, such an SPT-LSM constraint occurs precisely when $n\neq 0$. The generalization to arbitrary finite Abelian groups is straightforward (see Supplemental Material). %Appendix \ref{app: LSM}).

As an explicit example exhibiting an LSM constraint, consider arbitrary $N$ with $a=b=-1$ and $n=1$. We can write down a symmetric lattice Hamiltonian:
\begin{equation}\label{eq: Hal stagger}
    H=\sum_{i=1}^{L}\bigl(\cos\theta\, X_{i-1}X_i+\sin\theta\, Z_i Z_{i+1}\bigr)+\mathrm{h.c.},
\end{equation}
which preserves the exponential symmetries $\mathcal{U}_x=\bigotimes_{j=1}^L X_j^{(-1)^j}$ and $\mathcal{U}_z=\bigotimes_{j=1}^L Z_j^{(-1)^j}$. The LSM constraint then implies that this model cannot realize a unique gapped ground state; it must be either gapless or gapped with ground-state degeneracy. Interestingly, applying charge conjugation on the odd sites maps Eq.~\eqref{eq: Hal stagger} to the so-called quantum torus chain studied in Ref.~\cite{PhysRevB.86.134430,PhysRevB.104.045151}.

The map transforms the exponential symmetries into two unmodulated symmetries, which together with translation give rise to a conventional LSM theorem for the quantum torus chain.

In the second class, we consider a non-Abelian symmetry group for which only part of the symmetry is modulated. Specifically, we study a $\mathbb{Z}_N$ qudit chain with even $N$, whose symmetries are generated by
\begin{equation}\label{eq: D2N sym}
    \mathcal{U}_x=\bigotimes_{j=1}^L X_j^{a^{j-1}}, \qquad
    \mathcal{U}_c=\bigotimes_{j=1}^L C_j Z_j,
\end{equation}
where $a$ is coprime to $N$. Here $C_j$ denotes the charge-conjugation operator, acting as $C_j:X_j\mapsto X_j^\dagger$ and $Z_j\mapsto Z_j^\dagger$. Together, these symmetries generate the dihedral group $D_{2N}=\mathbb{Z}_N\rtimes\mathbb{Z}_2$. A nontrivial on-site projective structure is detected by the commutation relation between $\mathcal{U}_c$ and $X^{N/2}$, which exists only for even $N$ as $H^2(D_{2N},U(1))=\mathbb{Z}_{\gcd(N,2)}=\mathbb{Z}_2$. In the present case, the on-site projective class is characterized by the coboundary-invariant phase $\nu_j(x^{N/2},c)\nu_j(c,x^{N/2})^{-1}=(-1)^{a^{j-1}}=-1$, indicating that each site carries a nontrivial projective representation.

Assuming an injective MPS preserving these symmetries, the generalized push-through condition then gives
\begin{equation}
\begin{split}
  X_j^{a^{j-1}}\cdot A&\doteq(v_x^{\dagger})^{a^{j-1}} A v_x^{a^{j}}, \\
  CZ \cdot A&\doteq\, v_c^{\dagger} A v_c,
\end{split}
\end{equation}

where $v_x$ and $v_c$ act on the virtual indices. Since the $\mathbb{Z}_2$ subgroup has no nontrivial projective class by itself, we may take $v_c^2=\mathbbm{1}$. The projective representation on the virtual bond is characterized by $\omega_j(x^{N/2},c)\omega_j(c,x^{N/2})^{-1}
= v_x^{a^j N/2} v_c v_x^{-a^j N/2} v_c^{-1}$.
We assume the bond invariant at $j=0$ site is $\omega_0(x^{N/2},c)\omega_0(c,x^{N/2})^{-1}=(-1)^m$ with $m=0,1$. Then one finds $\omega_j(x^{N/2},c)\omega_j(c,x^{N/2})^{-1}=(-1)^{m a^j}$. Eq.~\eqref{eq:Grand_SPT_LSM intro} then requires $\omega_j(x^{N/2},c)\omega_j(c,x^{N/2})^{-1}
=\omega_{j-1}(x^{N/2},c)\omega_{j-1}(c,x^{N/2})^{-1}\nu_j(x^{N/2},c)\nu_j(c,x^{N/2})^{-1}$, which reduces to
\begin{equation}
m(a-1)=1 \pmod 2.
\end{equation}
This equation has no solution, since $a$ must be odd when it is coprime to even $N$. Therefore, no symmetric injective MPS exists, and an LSM obstruction follows.

The smallest nontrivial case is $N=4$. Taking $a=3\equiv -1 \pmod 4$, one may consider the following Hamiltonian
\begin{equation}
\begin{split}
    H= &-\sum_j X_j X_{j+1}^{\dagger}
    + J_1 \omega^{-1/2}X_j + \text{h.c.}\\
    &+ i J_2 X_j X_{j+1}
    + J_3 Z_j Z_{j+1}
    + \text{h.c.},
\end{split}
\end{equation}
with real couplings $J_1,J_2,J_3$. By the above analysis, this model cannot realize a unique gapped symmetric ground state. In particular, when $J_2=J_3=0$ and $J_1>0$, the model has two product-state ground states with $X_j=1$ and $X_j=i$, respectively, and thus spontaneously breaks the $\mathcal{U}_c$ symmetry. When $J_1=J_2=0$, the model also preserves the additional symmetry $\bigotimes_j Z_j$. In that limit, gaplessness is also enforced by the exponential LSM theorem of the first class, with $a=-1,\ b=1$ and $n=1$.

\newsec{Conclusion and discussion}
In this work, we developed an MPS framework for one-dimensional systems with modulated symmetries. Our main result is a generalized symmetry push-through condition: for an injective translationally invariant MPS, a modulated symmetry acting on the physical leg is absorbed into site-dependent unitaries on the virtual bonds. This extends the standard MPS description of global symmetries to the modulated setting and can be applied to SPT classification and LSM-type constraints. We illustrated the framework with exponential symmetries, mixed charge-exponential symmetries, and a non-Abelian example with partially modulated dihedral symmetry, recovering known SPT classifications, constructing explicit symmetric MPS representatives, and deriving lattice models with the predicted LSM obstructions.

Several important directions remain open. On the conceptual side, it would be valuable to extend the present framework beyond injective MPS to non-injective tensors, which can describe symmetry-breaking phases and other interesting long-range-entangled states with modulated symmetries. On the symmetry side, extending the formalism to continuous \cite{hu2024quantumbreakdownmodellattice,hu2025quantumbreakdowncondensatedisorderfree} and more general non-Abelian modulated symmetries would substantially broaden its scope. Finally, it would be particularly interesting to develop fermionic and higher-dimensional generalizations, for example by formulating analogous structures in fermionic tensor networks or in PEPS-like descriptions of modulated symmetry phases beyond one dimension.

\newsec{Acknowledgements} We thank Biao Lian for enlightening discussions on exponential symmetries. We also thank Sal Pace for discussions on the structure of modulated symmetries and for drawing our attention to prior work on the classification of modulated SPT phases. AA, SS, and ZB acknowledge support from NSF CAREER Grant No.~DMR-2339319. LHL and ZB also acknowledge partial support from a Quantum SuperSEED grant (ICDS\_QS25\_029093) from the Institute for Computational and Data Sciences at the Pennsylvania State University. 

\textit{Note added:} During the preparation of this manuscript, we became aware of an upcoming related work~\cite{lam-ning-2026}, which may have overlap with some of our results.

\bibliographystyle{unsrt} 
\bibliography{mspt}% Produces the bibliography via BibTeX.

\section{End Matter}
\newsec{Explicit MPS Tensor} Here we construct an explicit MPS representative for the family of exponential SPT phases with on-site group of $\mathbb{Z}_N\times \mathbb{Z}_N$, which satisfies Eq.~\eqref{eq:c-e constraint_main}. Consider an MPS tensor with physical Hilbert-space dimension $N^2$ and virtual bond dimension $N$, defined by $A^{s_1,s_2}=l_{s_1}^\dagger l_{s_2}$, where $l_s$ is the row vector $(l_s)_i=\delta_{si}$ and all indices range from $1$ to $N$. The global symmetries are generated by 
\begin{equation}
\mathcal{U}_{1} = \bigotimes_{j=1}^L U_{1}^{a^{j-1}},\qquad
\mathcal{U}_{2} = \bigotimes_{j=1}^L U_{2}^{b^{j-1}},
\end{equation}
with on-site actions
\begin{equation}
\label{eq:on-site-exponential_ab}
    \begin{split}
        \sum_{s_1',s_2'} U_1^{s_1s_2;\,s_1's_2'} A^{s_1',s_2'} &= \omega^{-k(s_1-a s_2)} A^{s_1,s_2},\\
    \sum_{s_1',s_2'} U_2^{s_1s_2;\,s_1's_2'} A^{s_1',s_2'} &= A^{s_1+1,s_2+b},
    \end{split}
\end{equation}
where $\omega=e^{2\pi i/N}$. A direct computation gives $U_1U_2U_1^\dagger U_2^\dagger=\omega^{k(ab-1)}$. Therefore, the on-site symmetry action furnishes a linear representation of $\mathbb{Z}_N\times\mathbb{Z}_N$ iff $\omega^{k(ab-1)}=1$, precisely reproducing the constraint that $k$ is a multiple of $N/\text{gcd}(ab-1,N)$. For $a=b$, this reproduces the result in Eq.~\eqref{eq: exp SPT constra}; for $a=1$, it reduces to the charge-exponential SPT case.

It is straightforward to verify that the on-site transformation rules in Eq.~\eqref{eq:on-site-exponential_ab} are generated by the bond-space actions $U_1\cdot A = (Z^k)^\dagger A (Z^{k})^a$ and $U_2\cdot A = X^\dagger A X^b$, where $Z$ and $X$ are the clock and shift operators for a $N$-dimensional qudit. Since $Z^k X = \omega^k X Z^k$, the bond unitaries furnish a projective representation of $\mathbb{Z}_N\times\mathbb{Z}_N$ labeled by the class $[k]\in H^2(\mathbb{Z}_N\times\mathbb{Z}_N,U(1))=\mathbb{Z}_N$, with $k$ being a multiple of $N/\text{gcd}(ab-1,N)$.

\newsec{More examples of edge state degeneracies} The model in Eq. \eqref{Eq:Exp_Chain_Numerical_Z8} belongs to a more general model:
\begin{widetext}
\begin{equation}\label{Eq:Exp_Chain_Numerical_END}
\begin{split}
    H= &-\sum_{j} \biggr( (\hat{L}_{j}^{+})^3\hat{s}^-_{j+1}  +h.c. \biggr) - J_1\sum_{j} \biggr( \hat{L}^-_j\hat{s}^+_j  +h.c. \biggr)-J_2\sum_j\biggr( \hat{s}^z_j + \hat{L}^z_j\biggr)^2 \\
    &- \delta \sum_{j} \biggr( \hat{s}^-_j\hat{L}^+_{j+1} + h.c.\biggr)- \eta \sum_{j} \biggr( \hat{s}^-_j\hat{L}^+_{j+2} + h.c.\biggr) - \zeta \sum_{j} \biggr( \hat{s}^-_j\hat{L}^+_{j+3} + h.c.\biggr).
\end{split}
\end{equation}
\end{widetext}
By tuning the parameters, this model can preserve different exponential symmetry and realizes the corresponding SPT phases.

\begin{figure}
\centering
\includegraphics[width=0.9\linewidth]{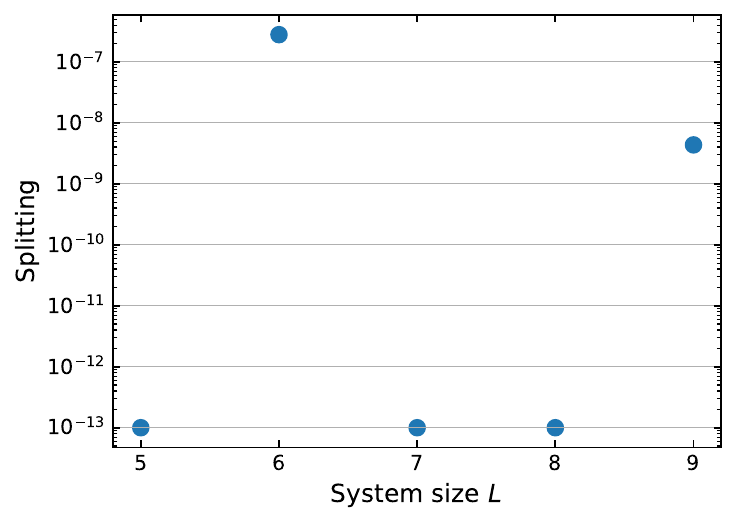}
\caption{Finite-size splitting of the edge states for the Hamiltonian in Eq.~\eqref{Eq:Exp_Chain_Numerical_END} with $J_1=J_2=0.4,\delta=\eta=0$ and $\zeta=0.5$, which corresponds to the case of $\mathbb{Z}_{26}$ exponential symmetry.}
\label{fig:numerics_APP}
\end{figure}

One may view the first line of Eq.~\eqref{Eq:Exp_Chain_Numerical_END} as a parent Hamiltonian for a $U(1)\rtimes\mathbb{Z}_2$ SPT with the exponentially modulated $U(1)$ generated by $\mathcal{U}_1(\theta)=\exp\{i\theta\sum_j 3^j(\hat{s}^z_j+\hat{L}^z_j)\}$ and a $\mathbb{Z}_2$, where the modulation base is (b=3), and the $\mathbb{Z}_2$ generated by spin-flip labeled by $\mathcal{U}_2$. The first term is an exact parent Hamiltonian for the nontrivial $U(1)\rtimes\mathbb{Z}_2$ SPT, whose fixed-point ground state can be represented as a MPS with bond dimension two. By contrast, the remaining terms in the first line are site-diagonal and favor a trivial product state. Their inclusion moves the system away from the fixed-point limit, rendering the Hamiltonian more generic and eliminating \emph{accidental} finite-size degeneracies. 

The perturbations ($\delta, \eta, \zeta$) in the second line carry exponential charges of $\pm 2,\pm 8,$ and $\pm 26$ respectively. Hence the $U(1)$ symmetry is broken down to $\mathbb{Z}_{2}, \mathbb{Z}_{8},$ or $\mathbb{Z}_{26}$. According to the analysis in the main text, exact edge-state degeneracies can occur only when both $b^L+1$ and $b^L-1$ are not divisible by the order of the residual symmetry group.

When $\delta\neq 0$ and $\eta=\zeta=0$, the $U(1)$ symmetry is broken down to $\mathbb{Z}_2$. Since the modulation base $b=3\equiv 1$ (mod $2$), this residual $\mathbb{Z}_2$ symmetry is effectively unmodulated. Thus there are no exact finite-size degeneracies remaining to be expected. 

\begin{figure}
\centering
\includegraphics[width=0.9\linewidth]{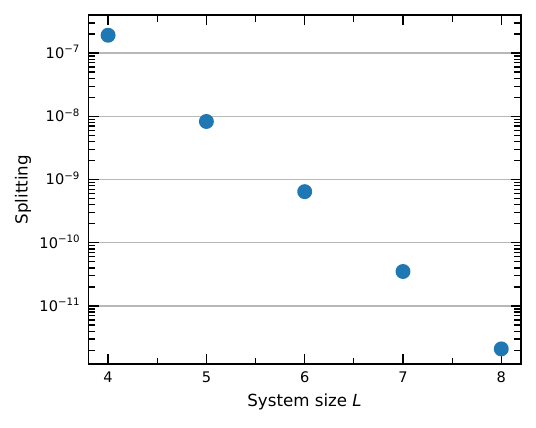}
\caption{Finite-size splitting of the %$\ket{\psi_{\alpha,\alpha}}$ multiplet of 
edge states for the Hamiltonian in Eq.~\eqref{Eq:Exp_Chain_Numerical_END} with $J_1=J_2=0.4,\zeta=\eta=0$ and $\delta=1$, which corresponds to the case of $\mathbb{Z}_{2}$ exponential symmetry.}
\label{fig:numerics_APP_Z2}
\end{figure}

When $\eta\neq 0$ and $\delta=\zeta=0$, the $U(1)$ symmetry is broken to $\mathbb{Z}_8$ and the model reduces to Eq.~\eqref{Eq:Exp_Chain_Numerical_Z8}. In this case, finite-size edge-state splitting occurs only when $3^L\pm 1=0\,(\text{mod } 8)$ which is satisfied only for even $L$.

Similarly, when $\zeta\neq 0$ and $\delta=\eta=0$, the $U(1)$ symmetry is broken to $\mathbb{Z}_{26}$. The condition for finite-size edge-state splitting becomes $3^L\pm 1=0\,(\text{mod } 26)$, which is satisfied only when $L=0\,(\text{mod } 3)$.

In Fig. \ref{fig:numerics_APP}, we compute the edge-state splitting using exact diagonalization for the case in where only $\zeta$ is nonzero, corresponding to an exponential symmetry with residual group $\mathbb{Z}_{26}$. We are able to verify that for small system sizes, finite-size splittings appear only at our predicted values of $L$ depending on the discrete symmetry group of the model. We can compare Figs. \ref{fig:main_splitting_fig}-\ref{fig:numerics_APP} to Fig. \ref{fig:numerics_APP_Z2}, with this latter figure showcasing the edge splittings for an unmodulated symmetry, which are nonzero for all $L$.

\end{document}

\title{Supplemental Materials --- Matrix Product States for Modulated Symmetries: SPT, LSM, and Beyond}
\author{Amogh Anakru}
\thanks{These authors contributed equally to this work.}
%\email{apa6106@psu.edu}
\affiliation{Department of Physics$,$ The Pennsylvania State University$,$ University Park$,$ Pennsylvania$,$ 16802$,$ USA}

\author{Linhao Li}
\thanks{These authors contributed equally to this work.}
\email{linhaoli601@gmail.com}
\affiliation{Department of Physics$,$ The Pennsylvania State University$,$ University Park$,$ Pennsylvania$,$ 16802$,$ USA}

\author{Sarvesh Srinivasan}
%\email{sbs5627@psu.edu}
\affiliation{Department of Physics$,$ The Pennsylvania State University$,$ University Park$,$ Pennsylvania$,$ 16802$,$ USA}
\author{Zhen Bi}
\email{zjb5184@psu.edu}
\affiliation{Department of Physics$,$ The Pennsylvania State University$,$ University Park$,$ Pennsylvania$,$ 16802$,$ USA}
\affiliation{Center for Theory of Emergent Quantum Matter$,$ Institute for Computational and Data Sciences$,$ The Pennsylvania State University$,$ University Park$,$ Pennsylvania$,$ 16802$,$ USA}
\date{\today}
	\maketitle

\section{Review of Matrix Product State Formalism}

Here, we briefly review key aspects of matrix product states (MPS) that are essential for understanding the present work. We point the reader to Refs.~\cite{perez2006matrix, pedagogical_pushthrough,cirac_review} for a more complete review of MPS.

We focus on translationally invariant matrix product states generated by the three-legged tensor $A^s$ depicted in Eq.~\eqref{eq:MPS_intro}; here the index $s$ describes the vertical (physical) leg and then $A^s$ is then a matrix in bond space (indicated by horizontal legs). A state $\ket{\psi}$ of length $L$ with periodic boundary conditions can then be constructed as follows:

\ie\label{eq:MPS_intro}
\vcenter{\hbox{\includegraphics[scale=0.25]{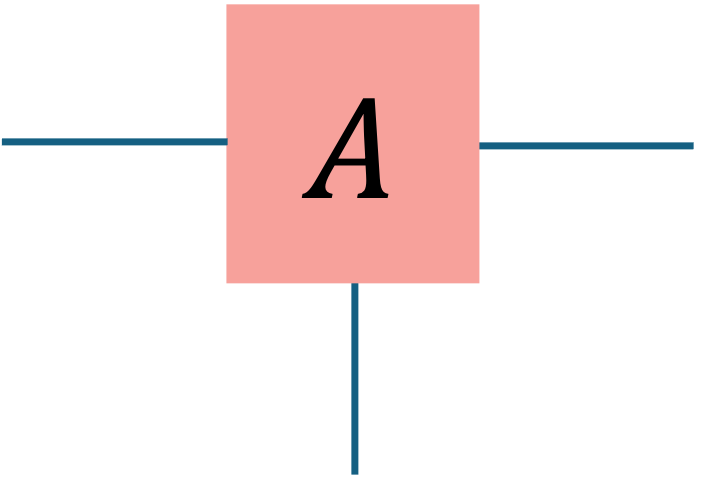}}} \ \implies \ \vcenter{\hbox{\includegraphics[scale=0.25]{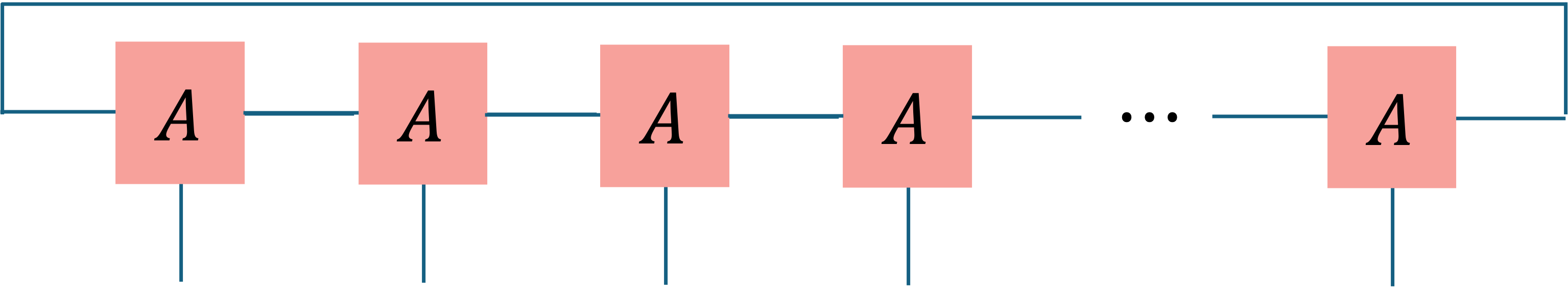}}} 
\fe

which represents the state
\begin{equation}
    \ket{\psi} = \sum_{s_1,\ldots,s_L} \Tr(A^{s_1}\ldots A^{s_L}) \ket{s_1\ldots s_L}.
\end{equation}

We also define a \textit{transfer matrix} associated with the MPS tensor:
\ie\label{eq:MPS_transfer}
\mathbb{E} =  \ \vcenter{\hbox{\includegraphics[scale=0.25]{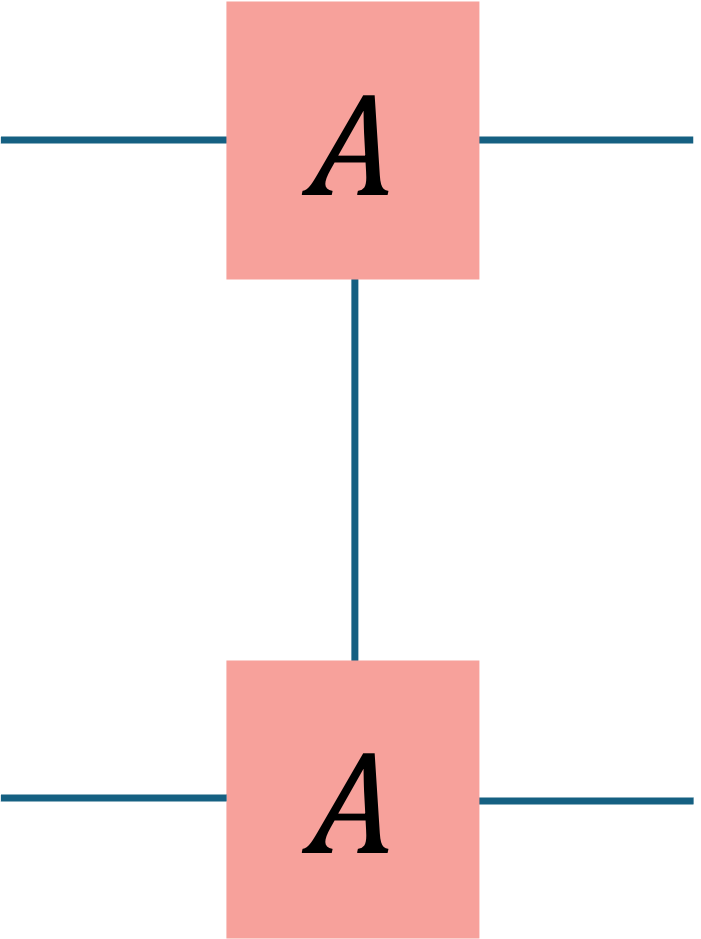}}},
\fe
where we work with a convention where the MPS tensor $A$ with physical leg pointing upward is understood to be complex-conjugated. Multiple different tensors $A^{s}$ can correspond to the same physical periodic state. One usually works with the left-canonical form defined by the following conditions:

\ie\label{eq:canonical}
\vcenter{\hbox{\includegraphics[scale=0.25]{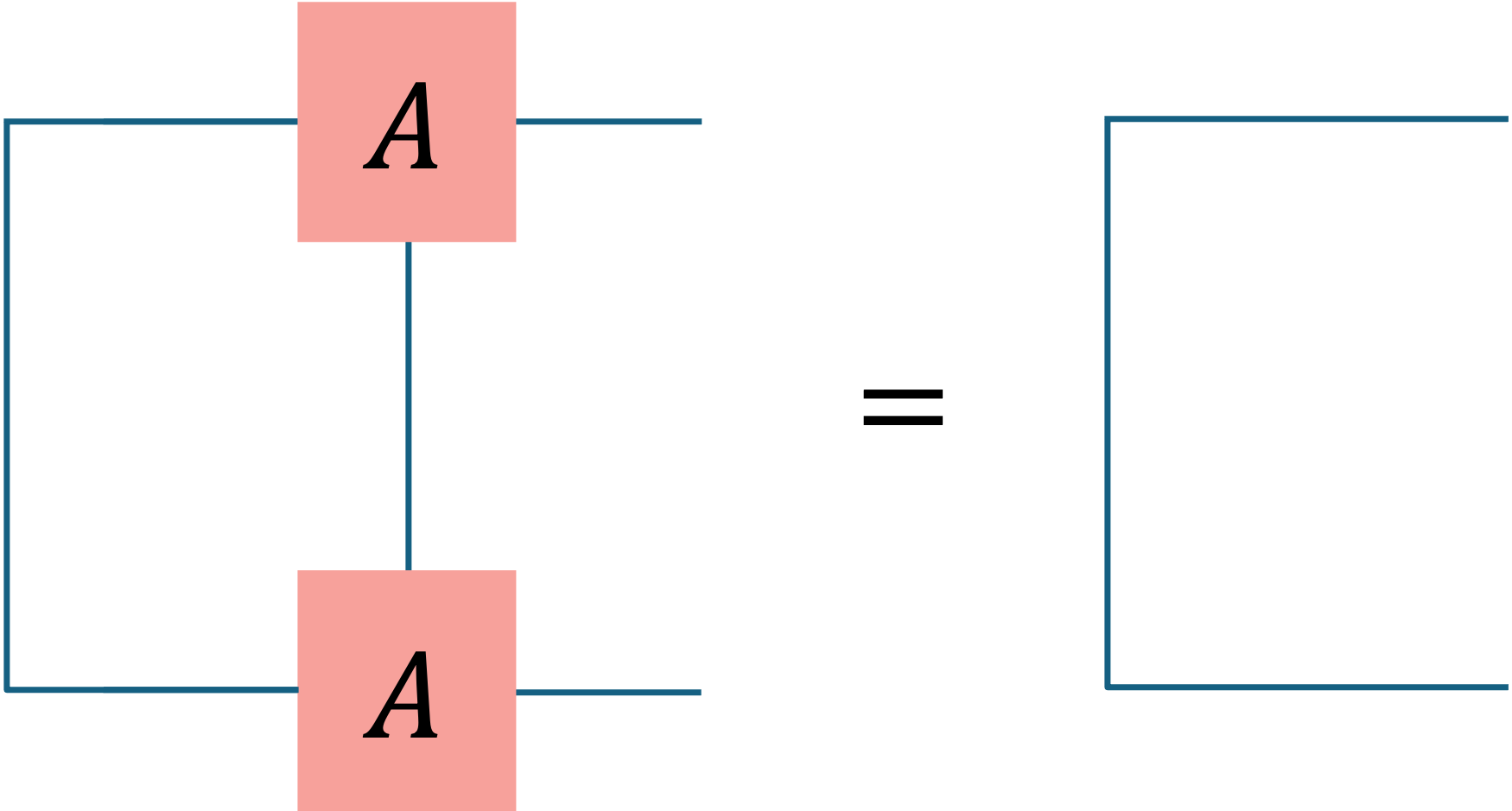}}}\ \,\,\,\, ,  \  \vcenter{\hbox{\includegraphics[scale=0.25]{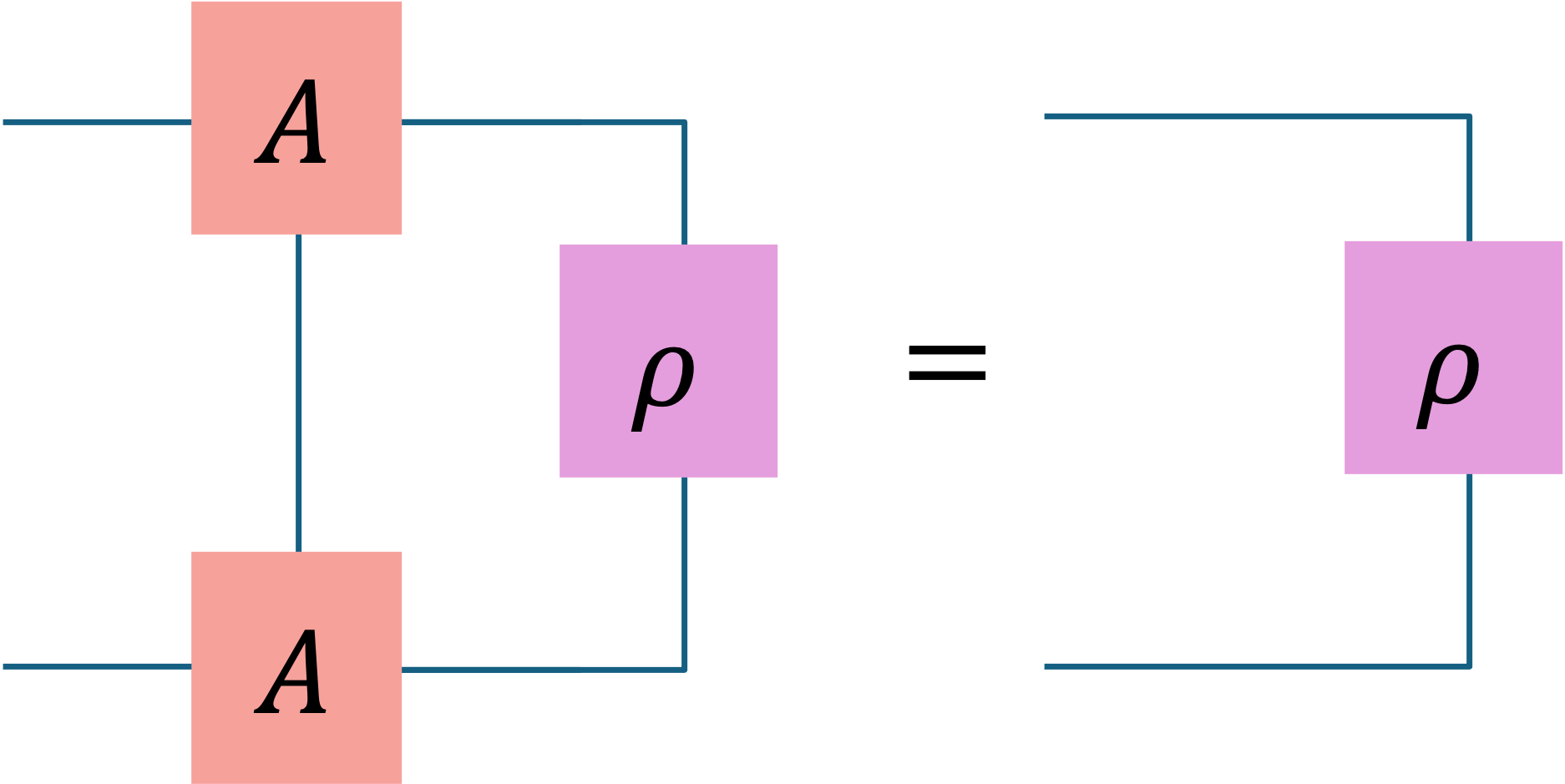}}}
\fe
with these left and right eigenvectors of the transfer matrix being dominant eigenvectors. Any translationally invariant MPS can be transformed into this form without changing the physical state. The matrix $\rho$ is full-rank and positive definite. The MPS is called \textit{injective} if these dominant eigenvectors are unique. A consequence of injectivity is that there is a finite `injective length' $\ell_0$ such that for $\ell\geq \ell_0$, the matrices $A^{s_1}\cdots A^{s_{\ell}}$ span the set of all matrices in bond space (these become matrices after fixing the indices $s_i$). \hfill \break

A final property is that of \textit{parent Hamiltonians}. An injective translationally-invariant MPS admits a Hamiltonian whose unique gapped ground state is the MPS. The Hamiltonian is constructed as follows:
\begin{equation}\label{eq:Master_Parent_Hamiltonian}
    \hat{H} = \sum_{j=1}^L \mathbb{I} - \mathcal{P}_{j,\ldots,j+\ell}
\end{equation}
where $\ell \geq \ell_0$ and $d^\ell> \chi^2$ where $d,\chi$ are respectively the on-site and bond dimensions. Define a map

\begin{equation}\label{eq:bond_physical_map}
\Gamma_j: M\mapsto \sum_{s_j\ldots s_{j+\ell}}\Tr(MA^{s_j}\cdots A^{s_{j+\ell}})\ket{s_{j}\ldots s_{j+\ell}}.
\end{equation}
The projector $\mathcal{P}_{j,\ldots,j+\ell}$ is the projector into the image of $\Gamma_j$. Since $\ell \geq \ell_0$, the Hamiltonian annihilates the MPS due to the injectivity condition, and as $d^\ell > \chi^2$, the projector has a nontrivial kernel and thus $\hat{H} \neq 0$. Accordingly, the injective MPS representatives constructed in the main text give rise to explicit parent Hamiltonians for SPT phases.

\section{Generalized push-through theorem}\label{app:pushthrough_proof}

We consider a left-canonical, injective MPS tensor $A^{s}$ which defines a translationally invariant periodic wavefunction $\ket{\psi}$ on $L$ sites that is invariant under certain modulated symmetry $\mathcal{U}_g=\bigotimes_{j=1}^{L} U_{g,j}$. To obtain a well-defined thermodynamic limit, we further assume that for every positive integer $k$, the MPS on a chain of length $kL$ (formed by concatenating $k$ identical $L$-site unit cells) is also an eigenstate of the $k$-fold periodic extension of the symmetry $ \bigotimes^k\mathcal{U}_g$. This ensures that the modulated symmetry structure extends consistently to arbitrary large system size. 
%We refer to this as the \emph{unravelable} condition for the $L$-site MPS.

In this setting, we now prove the key result: on each site we have the symmetry push-through condition 

\ie\label{eq:Grand_Pushthrough_App}
\vcenter{\hbox{\includegraphics[scale=0.35]{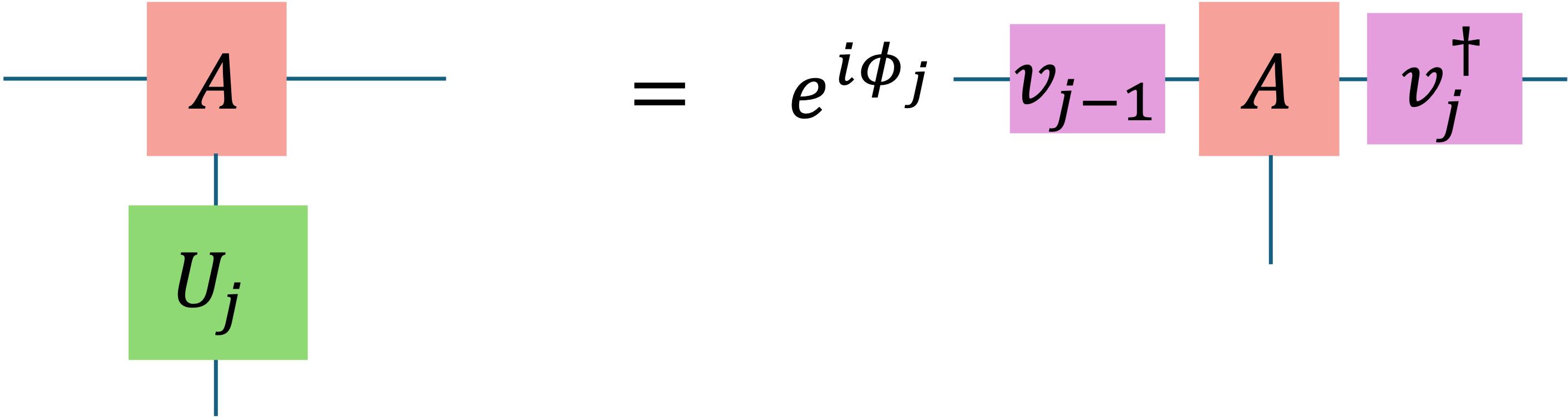}}}
\fe
with $v_j$ unitary (here and in what follows, we suppress the group label whenever no confusion can arise). Note that the convention of which bond is conjugated has changed from the main text. We note that if this condition is correct, it is trivial to see that the MPS is symmetric under the modulated symmetry. 

\subsection{Proof strategy}
For unmodulated symmetries, the conventional push-through rule is $U\cdot A = e^{i\phi}vAv^\dag$. This is proven \cite{cirac_pushthrough,pedagogical_pushthrough} by defining a structure known as a $U$-transfer matrix $\mathbb{E}_U$, given by the tensor contraction
\ie\label{eq:U-transfer-matrix}
\mathbb{E}_U = \vcenter{\hbox{\includegraphics[scale=0.25]{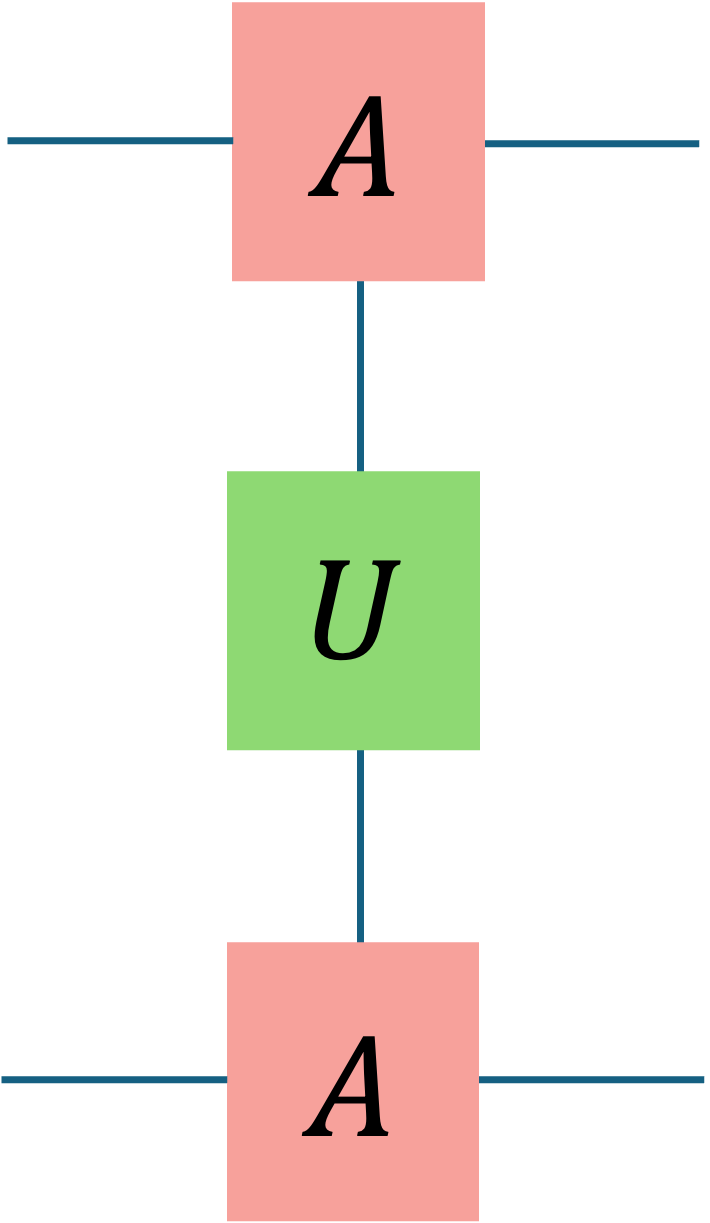}}}.
\fe
The total symmetry operator is $\mathcal{U}= U^{\bigotimes L}$. One can show that the expectation $\abs{\bra{\psi}\mathcal{U}\ket{\psi}}=1$ iff the $U$-transfer matrix has unit modulus eigenvalue. The consequence of this is then the push-through condition. When the on-site unitary is a dipole \cite{ho_tat_lam_dipolar_SPT_classification} or multipole symmetry \cite{Saito_Multipolar_MPS}, a similar proof strategy is possible by defining suitable generalizations of a $U$-transfer matrix which are forced to have a dominant eigenvalue on the unit circle, because the expectation $\abs{\bra{\psi}\mathcal{U}\ket{\psi}}$ turns out to be expressible as $\Tr(\tilde{\mathbb{E}}^L)$ for a suitable redefined $\tilde{\mathbb{E}}$.

For general modulated symmetry, denoting the now site-dependent $U$-transfer matrix as $\mathbb{E}_{U_{(j)}}$, it is not clear at all that $\abs{\bra{\psi}\mathcal{U}\ket{\psi}}$ is expressible as $\Tr(\tilde{\mathbb{E}}^L)$. Thus, we need to demand a different condition. We note that one proof strategy for general modulated symmetries involves invoking a Fundamental Theorem for injective MPS \cite{Molnar_2018}, which was used in Ref.~\cite{stephen2019subsystem} to prove the result. We will present a different proof that directly generalizes the strategies of \cite{pedagogical_pushthrough,ho_tat_lam_dipolar_SPT_classification, Saito_Multipolar_MPS}, and also allows for slight generalizations of the push-through result as stated in \cite{stephen2019subsystem}.

\subsection{Proof for generalized push-through condition}

We will start off by presenting some generalizations of results proven in \cite{pedagogical_pushthrough, cirac_pushthrough}. These results are concerned with the "$U$-transfer matrix" defined in Eq.~\eqref{eq:U-transfer-matrix}.

We first prove two lemmata:
\begin{lemma}\label{lem:1}
Let $U$ be an on-site unitary, and consider the following diagrammatic equation: 
\ie\label{eq:Generalized_Transfer_Condition}
\vcenter{\hbox{\includegraphics[scale=0.25]{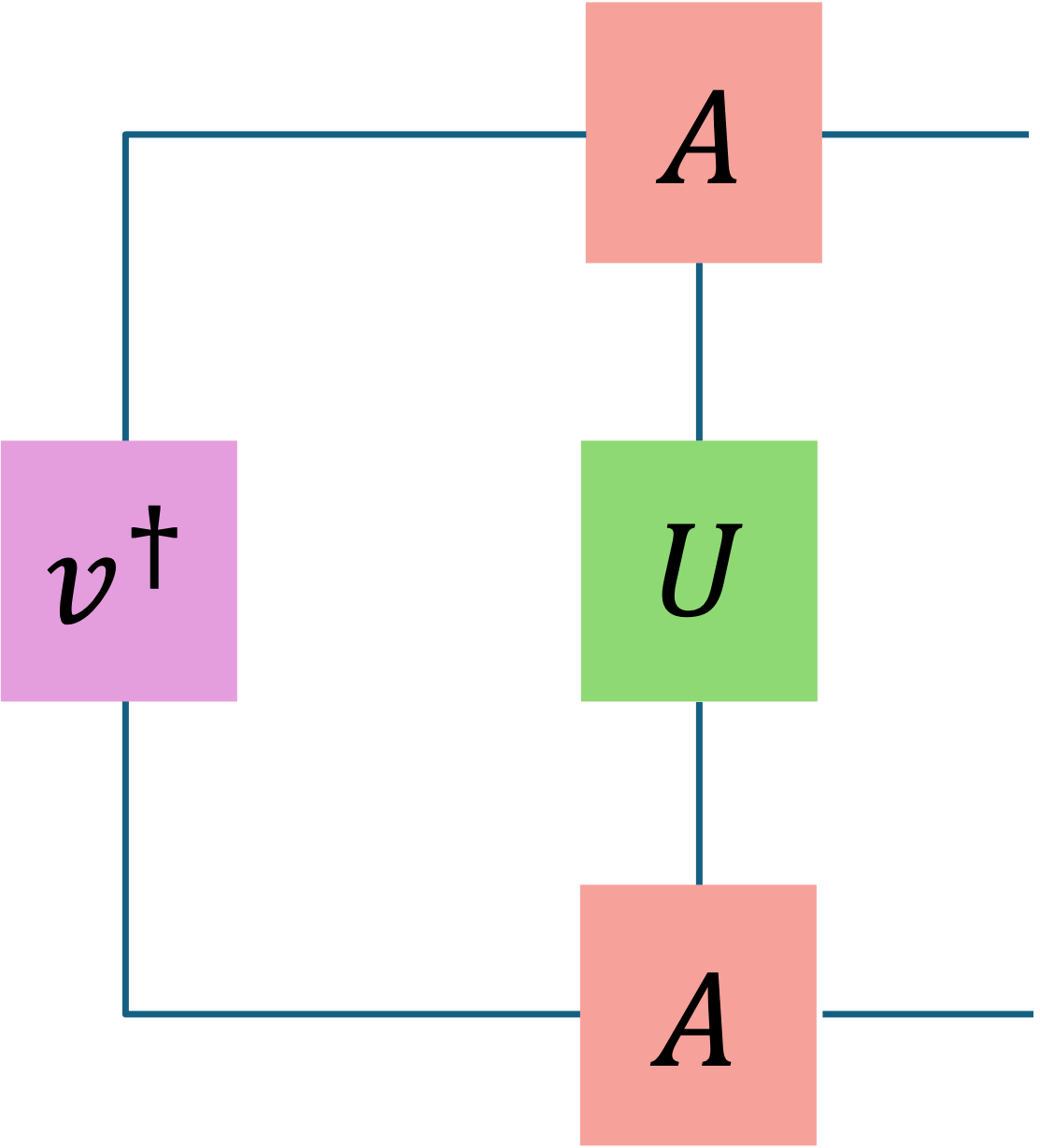}}}\ = \lambda\ \vcenter{\hbox{\includegraphics[scale=0.25]{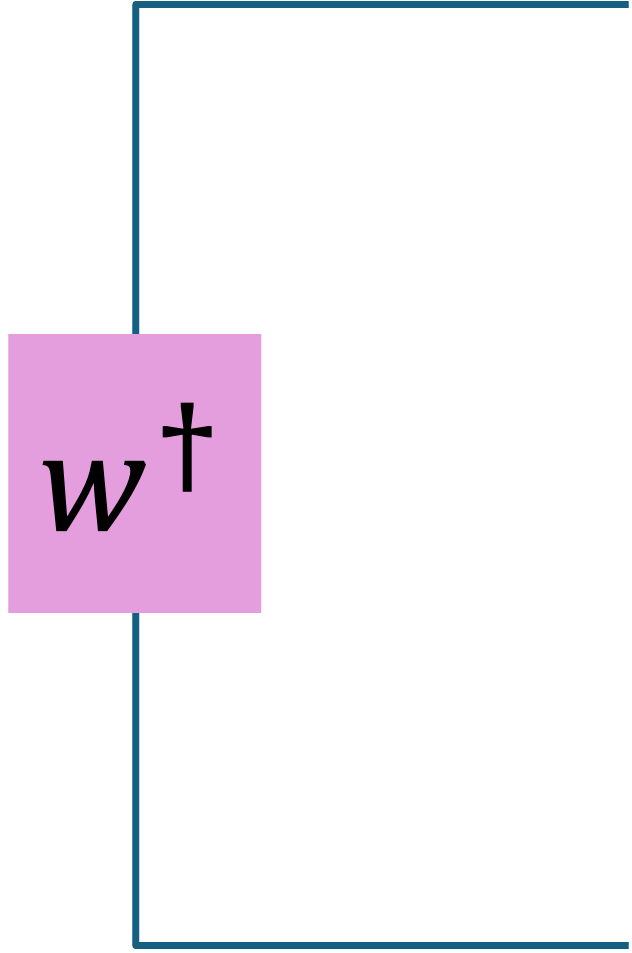}}}
\fe
with normalization conditions $\Tr(v^\dag v \rho)= \Tr(w^\dag w \rho)=1$. Here $\rho$ is the unique dominant right eigenvector of the transfer operator associated with the injective MPS $A$. Then we have $\abs{\lambda}\leq 1$. \hfill \break
\end{lemma}

\begin{remark*}
$\rho$ is full-rank and $v^\dag v, w^\dag w$ are positive semi-definite and nonzero. Thus the condition on $v,w$ is a valid normalization choice. \hfill \break
\end{remark*}

\begin{proof}[\emph{\textbf{Proof:}}]
 First, we define an inner product on bond space by 

\ie\label{eq:generalized_inner_product}
\langle v^\dag,w^\dag \rangle_T = \Tr(w^\dag v \rho) = \vcenter{\hbox{\includegraphics[scale=0.25]{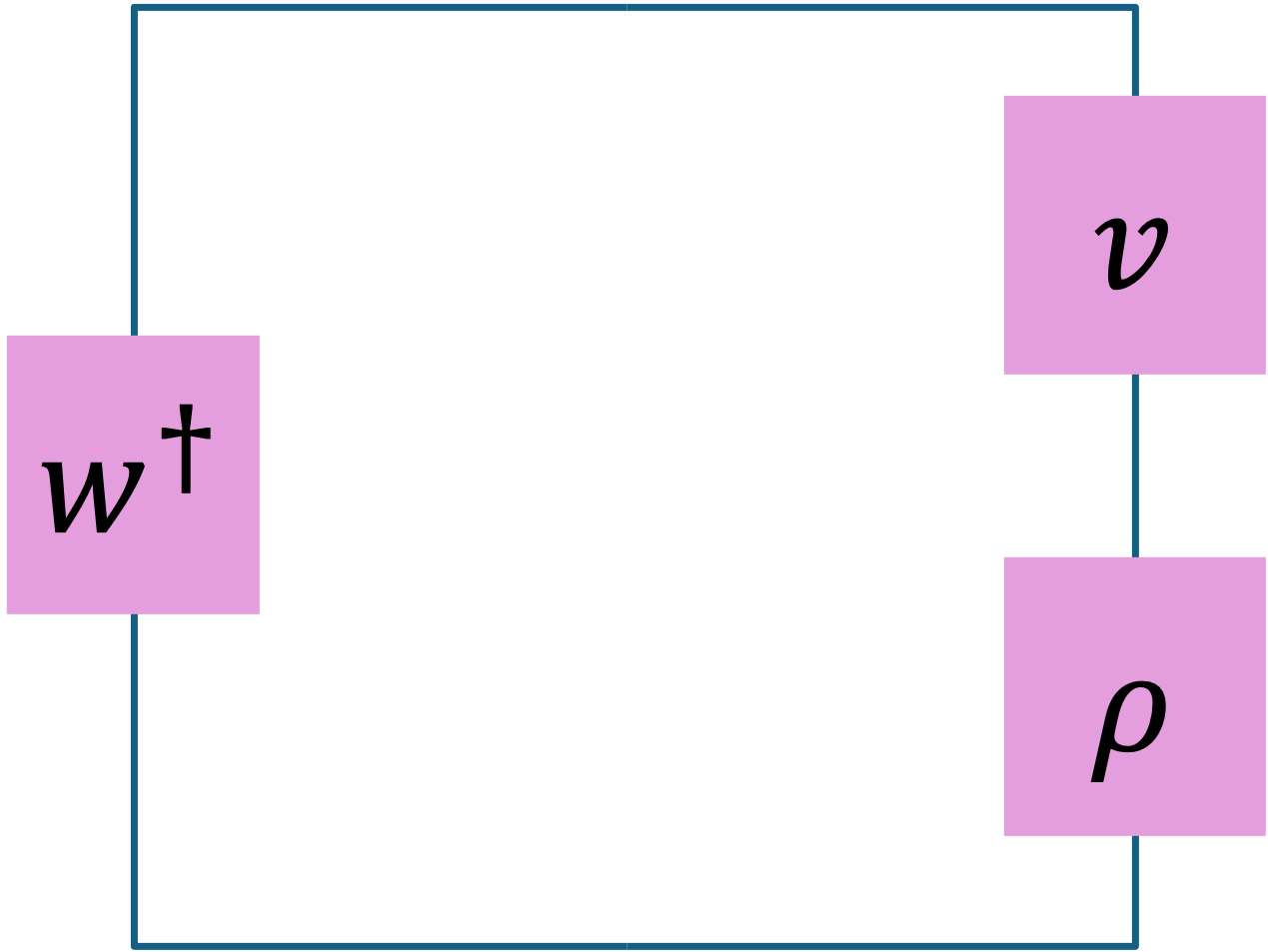}}}
\fe
and define an operator norm $\mathnorm{\cdot}$ induced by the inner product $\langle\cdot,\cdot\rangle_T$, then Lemma \ref{lem:1} is equivalent to showing that we have $\mathnorm{\mathbb{E}_U} \leq 1$. It is easy to verify that Eq.~\eqref{eq:generalized_inner_product} defines an inner product. The nontrivial properties worth checking are conjugate symmetry\footnote{We have $\langle v^\dag,w^\dag \rangle_T^* = (\Tr(w^\dag v \rho))^\dag = \Tr(\rho v^\dag w) = \langle w^\dag,v^\dag \rangle_T$ by cyclicity of trace.} and positive definiteness\footnote{$\langle v^\dag,v^\dag \rangle_T = \Tr(v\rho v^\dag) = \sum_{\alpha}(\bra{\alpha}v\sqrt{\rho})(\sqrt{\rho}v^\dag\ket{\alpha})$. The summands are all nonnegative and if every summand is zero, this implies that $\sqrt{\rho}v^\dag = 0 \implies v^\dag=0$ since $\rho$ is full-rank. Thus $\langle v^\dag,v^\dag \rangle_T>0$ for $v^\dag\neq 0$.}. We now use this norm to prove the lemma. We compute:

%\begin{equation}\label{eq:Cauchy-Schwarz_Horizontal}
%    \abs{\lambda}^2\abs{\Tr(w^\dag w \rho)}^2 = \abs{\mathcal{E}_U(v^\dag)w\rho}^2 = \biggr\lvert\biggr\langle \sqrt{U^\dag}v A\sqrt{\rho}   ,\,\, \sqrt{U}Aw\sqrt{\rho}\biggr\rangle\biggr\rvert^2 \leq \Tr(w^\dag w \rho) \Tr(v^\dag v \rho)
%\end{equation}

\ie\label{eq:cauchy_schwarz_steps_1}
\ \abs{\lambda}^2\abs{\langle w^\dag,w^\dag\rangle_T}^2  \ =\ \vcenter{\hbox{\includegraphics[scale=0.25]{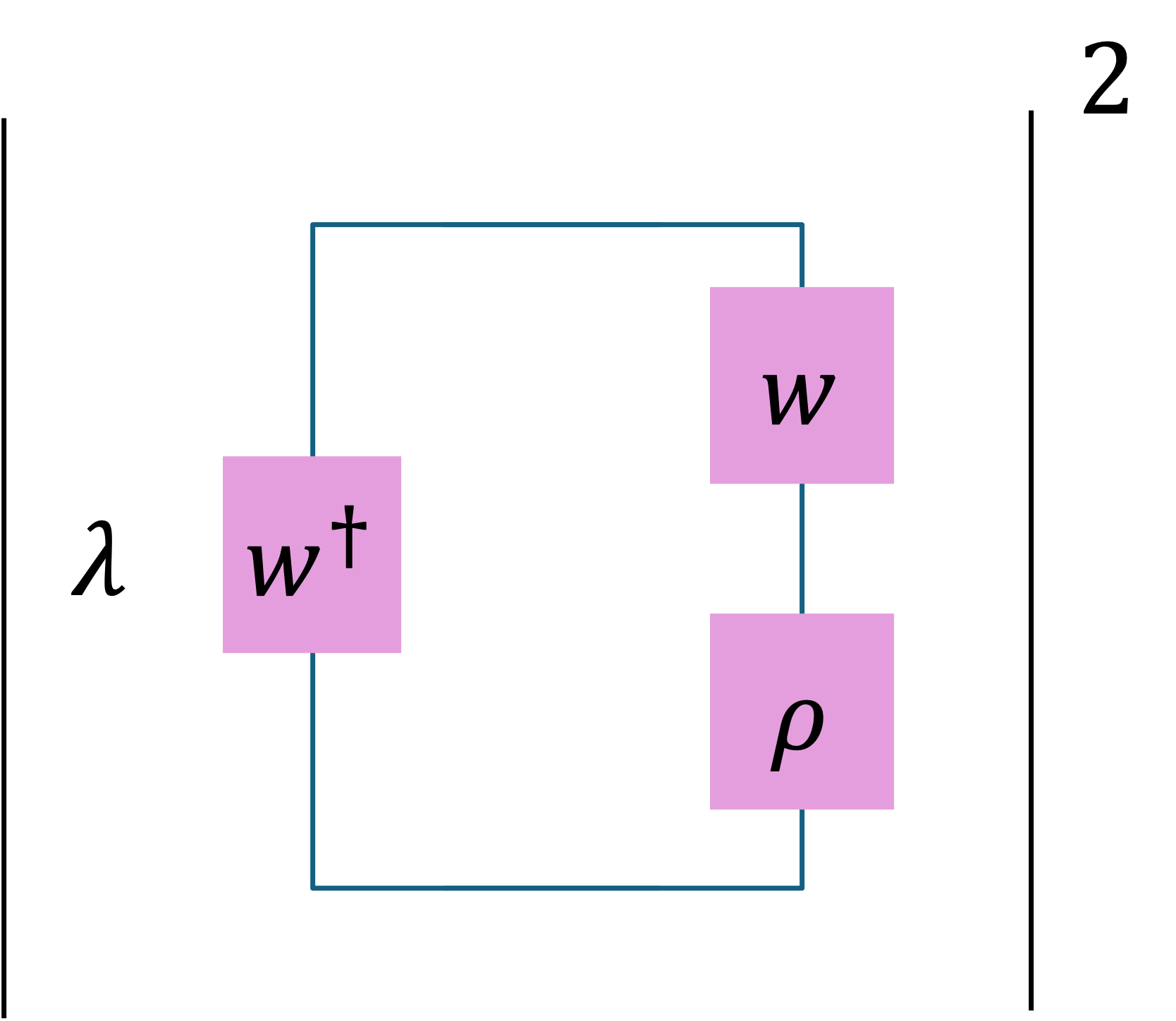}}}  \ = \ \vcenter{\hbox{\includegraphics[scale=0.25]{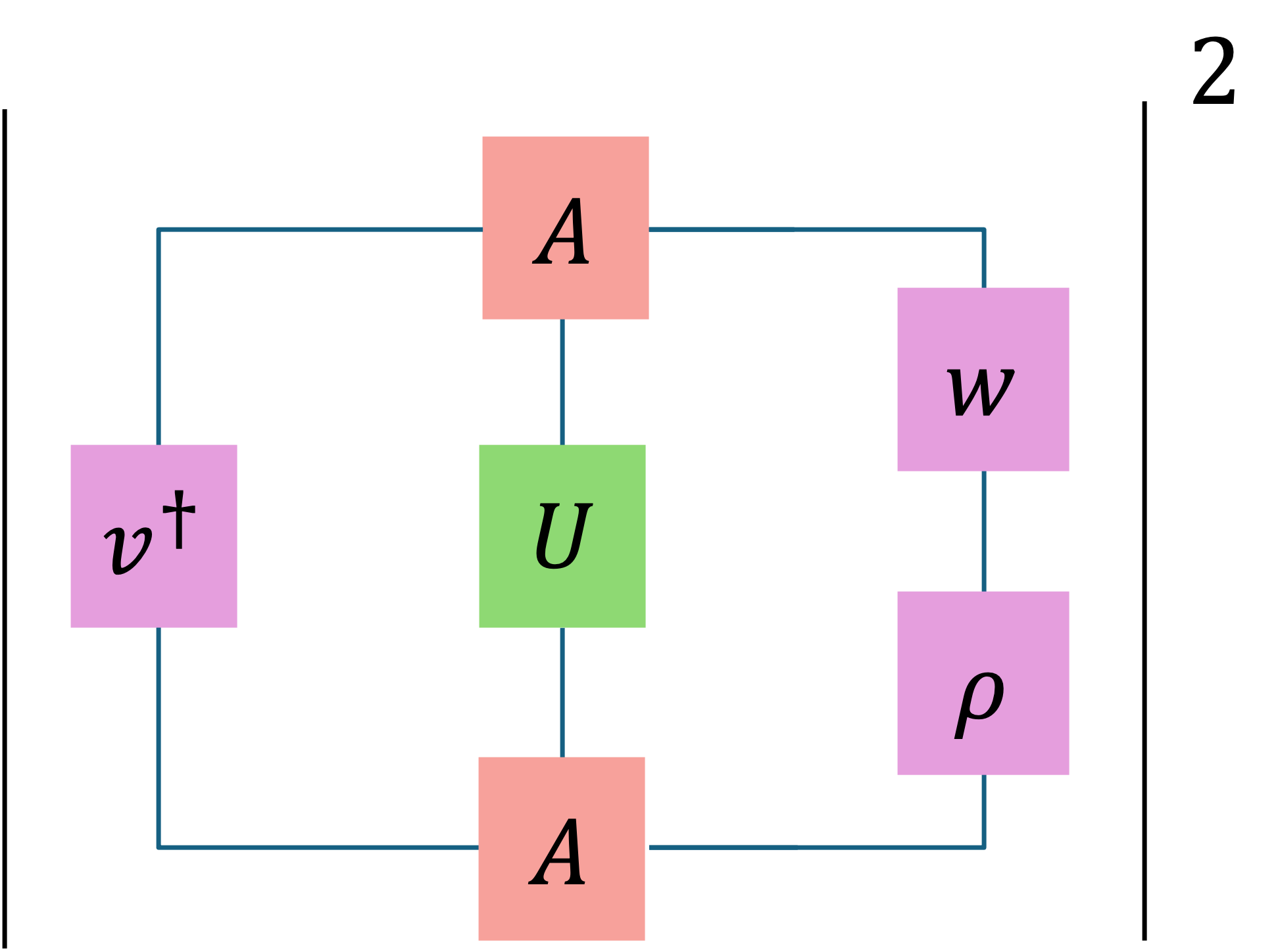}}} \ = \ \vcenter{\hbox{\includegraphics[scale=0.25]{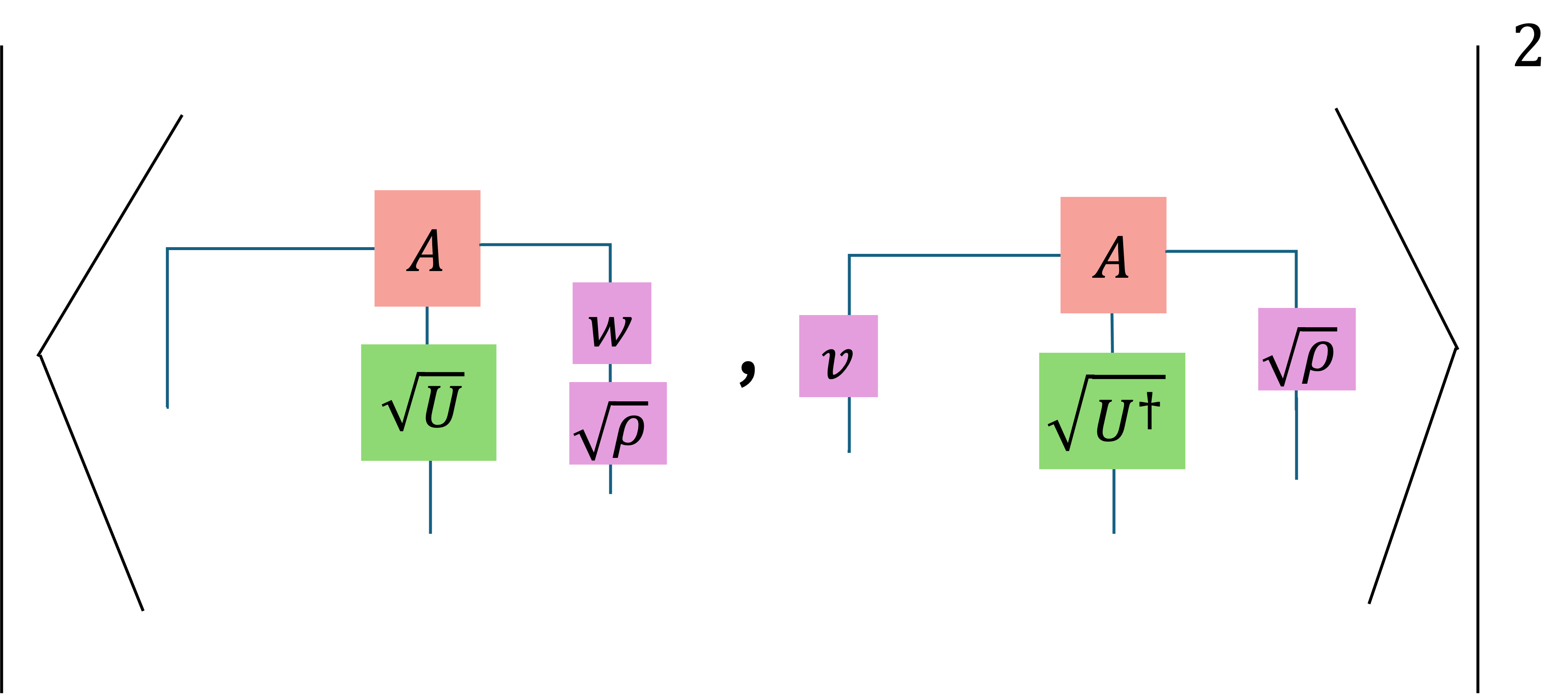}}}
\fe

\ie\label{eq:cauchy_schwarz_steps_2}
\ \leq\ \vcenter{\hbox{\includegraphics[scale=0.25]{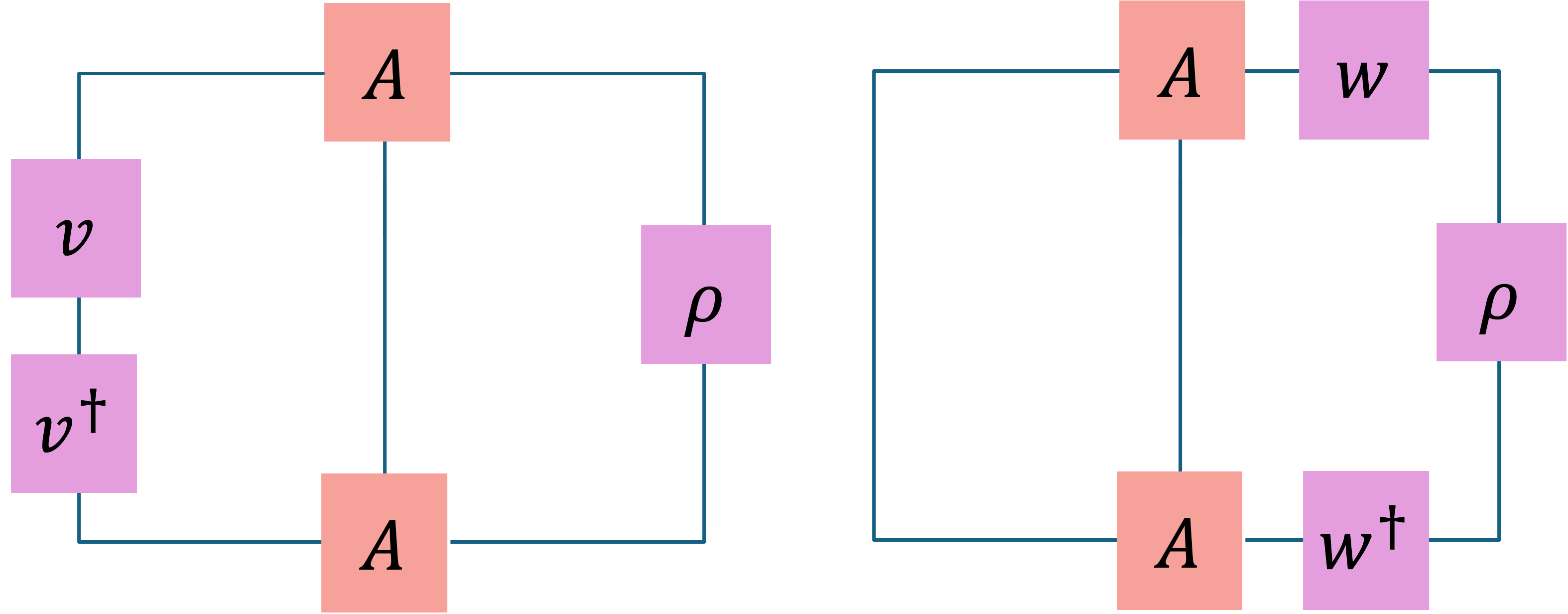}}} \ = \ \vcenter{\hbox{\includegraphics[scale=0.25]{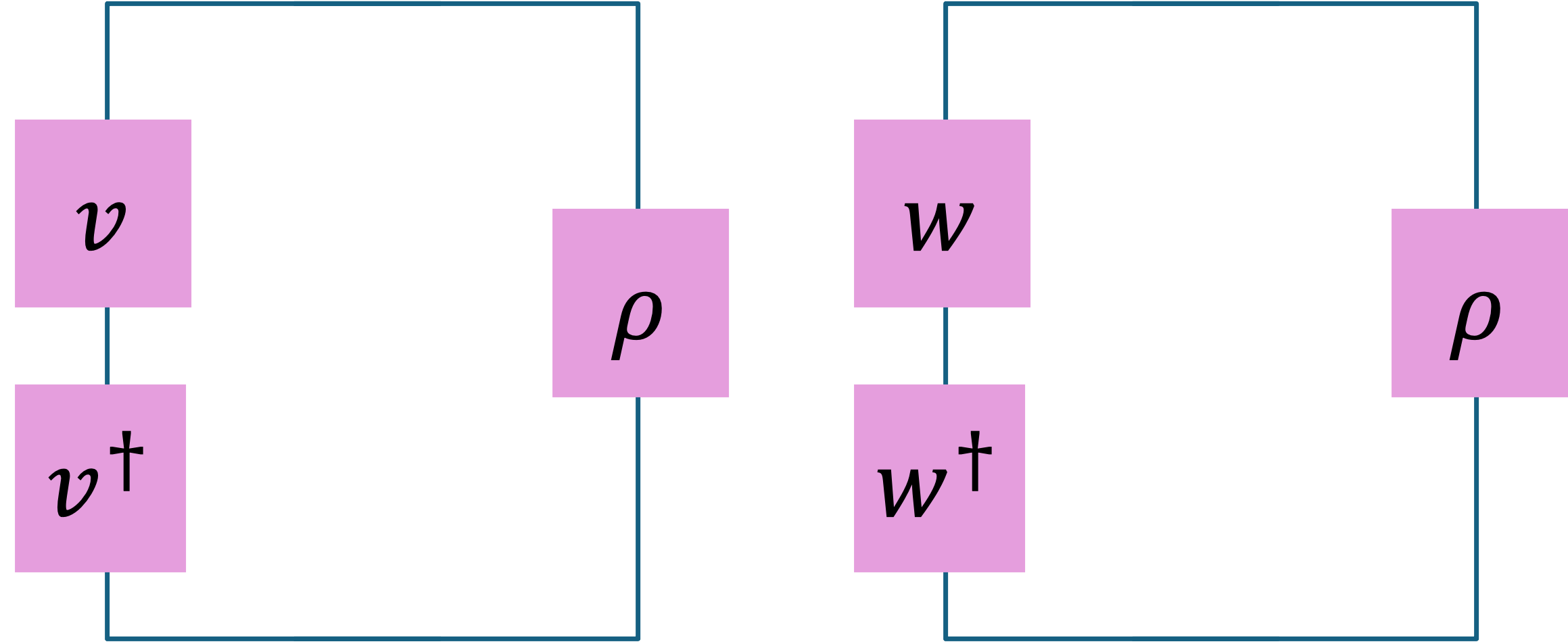}}} \ = \langle v^\dag,v^\dag\rangle_T \cdot \langle w^\dag,w^\dag\rangle_T .
\fe
In order, each (in)equality follows from: 1) Definition \eqref{eq:generalized_inner_product} ; 2) Eq.~\eqref{eq:Generalized_Transfer_Condition}; 3) interpretation as a $\mathbb{C}^N$ inner product of vectors; 4) Cauchy-Schwarz; 5) canonical form and the fact that $\mathbbm{1}, \rho$ are left/right dominant eigenvectors; 6) Definition \eqref{eq:generalized_inner_product}. The lemma follows once we take $\langle v^\dag,v^\dag\rangle_T = \langle w^\dag,w^\dag\rangle_T =1$ per the assumption. It is also clear why we needed to define this unconventional norm in order to get a concrete bound on $\abs{\lambda}$. \hfill \break

\end{proof}

\begin{lemma}\label{lem:2}
    The inequality $\abs{\lambda}\leq 1$ becomes an equality if and only if we have a "weak push-through" condition

\ie\label{eq:weak_pushthrough}
\vcenter{\hbox{\includegraphics[scale=0.3]{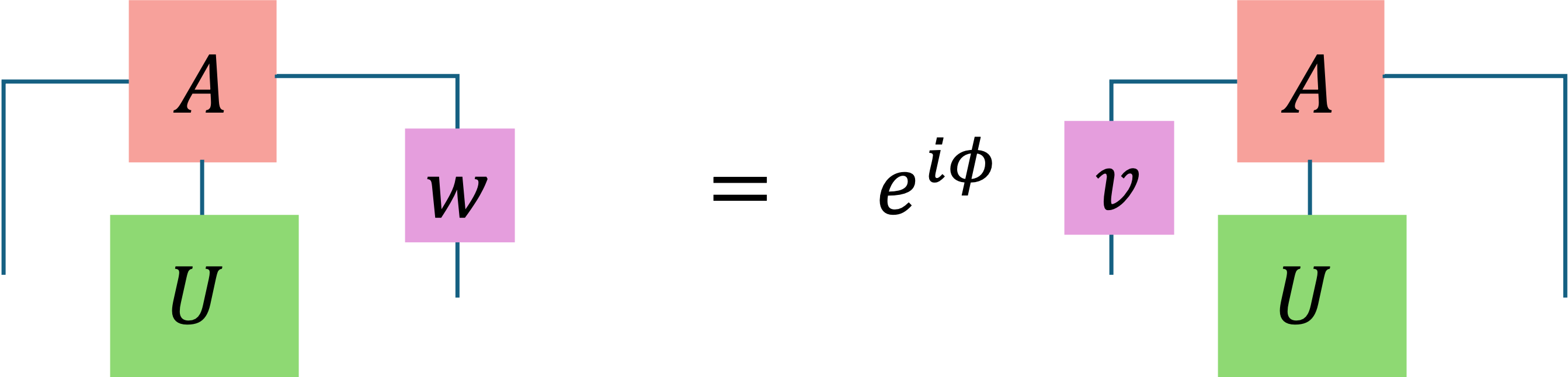}}}
\fe
and furthermore that
\ie\label{eq:near_transference}
\vcenter{\hbox{\includegraphics[scale=0.25]{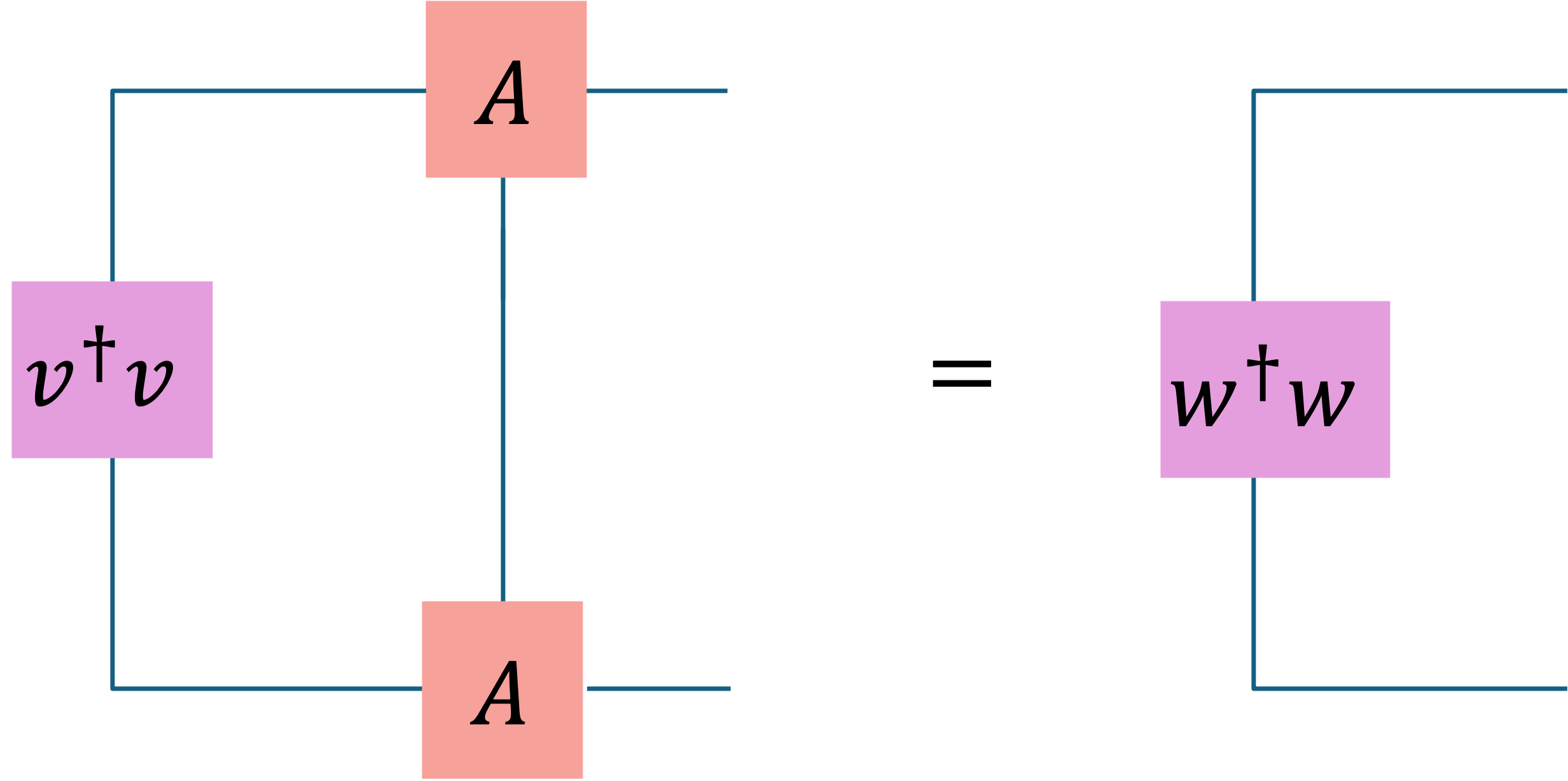}}} \ \implies \mathcal{E}(v^\dag v) \,{\doteq}\, w^\dag w,
\fe
where we have written the action of the original transfer matrix $\mathbb{E}$ as a channel $\mathcal{E}(\cdot)$ acting on matrices.

\end{lemma}
\begin{proof}[\emph{\textbf{Proof:}}]
The if-direction is simple. To get the only-if direction, we suppose $\abs{\lambda}=1$. Then, the Cauchy-Schwarz inequality between Eqs. \eqref{eq:cauchy_schwarz_steps_1} and \eqref{eq:cauchy_schwarz_steps_2} will be an \textbf{equality}. Note that the two vectors in the inner product in the last line of Eq.~\eqref{eq:cauchy_schwarz_steps_1} have the same norm, thus they differ only by a phase. Inverting $\rho$ and $\sqrt{U^\dag}$, Eq.~\eqref{eq:weak_pushthrough} follows with $\lambda = e^{i\phi}$. At the present stage, we have \textbf{not} used or proved that $v,w$ are unitary or even invertible. \hfill \break 

Now we need to prove the second half of the lemma that $\mathcal{E}(v^\dag v) \,{=}\, w^\dag w$. We start with Eq.~\eqref{eq:Generalized_Transfer_Condition}, and by assumption, we have $\lambda=e^{i\phi}$. Now multiply the top leg of both sides by $w$:

\ie\label{eq:lemma_2_second_half}
\vcenter{\hbox{\includegraphics[scale=0.25]{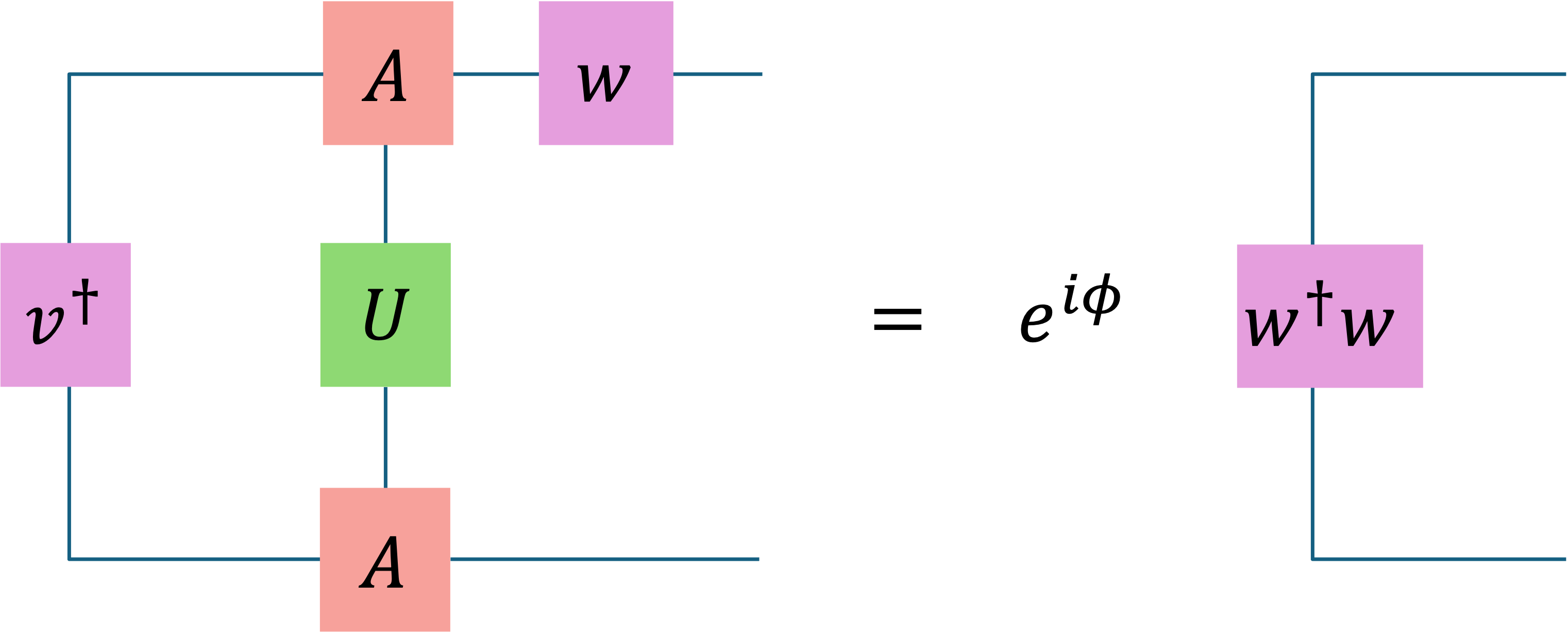}}}. 
\fe

Then Eq.~\eqref{eq:near_transference} follows from a simple application of Eq.~\eqref{eq:weak_pushthrough}.

\end{proof}
\begin{remark*}
   If $v^\dag v = w^\dag w$, injectivity would immediately imply that $v,w$ are unitary. However, it does \textit{not} follow from what we have shown so far that $v^\dag v = w^\dag w$. As an example, let $U=1$ so that we are discussing just the original transfer operator $\mathcal{E}$. Consider an eigenvalue/eigenvector pair $\mathcal{E}(X) = \lambda X$ with $\abs{\lambda}< 1$. (Note that we also have $\mathcal{E}(X^\dag) = \lambda^* X^\dag$.) We can choose $\gamma$ sufficiently small such that $Y = \mathbbm{1}+\gamma(X+X^\dag)$ is positive definite and Hermitian and thus can be written as $v^\dag v$. Of course $\mathcal{E}(Y)=\mathcal{E}(v^\dag v)=w^\dag w$ as this is also positive-definite. Furthermore, left/right orthonormality of eigenvectors guarantees that $v^\dag$ and $w^\dag$ have the same norm (as defined by the inner product in Eq.~\eqref{eq:generalized_inner_product}). Yet, $v^\dag v \neq w^\dag w$.  
\end{remark*}

\begin{lemma}\label{lem:3}
If a product of $U$-transfer matrices $\mathbb{E}_U^{1}\dots \mathbb{E}_U^{n}$ has operator norm equal to $1$, then each $\mathbb{E}_U^j$ has $v^\dag_j,w^\dag_j$ satisfying Eq.~\eqref{eq:Generalized_Transfer_Condition} with $\abs{\lambda}=1$ , $w_{n-1}^\dag \dot{=} v_n^\dag$, and thus $w_n^\dag w_n = \mathcal{E}(w_{n-1}^\dag w_{n-1})$. (We define $w_0^\dag = v^\dag$.) \end{lemma}
\begin{proof}[\emph{\textbf{Proof:}}]This is a straightforward application of the inequality $\mathnorm{\mathbb{E}_U^{1}\dots \mathbb{E}_U^{n}} \leq \prod_{j=1}^n\mathnorm{\mathbb{E}_U^{j}} \leq 1$ (here the latter inequality comes from Lemma \ref{lem:1}), as well as Lemma \ref{lem:2}. In other words, each application of an $\mathbb{E}^j_U$ either preserves or decreases the norm of the vector it acts on. Note that we have not assumed that such a choice of $v_j,w_j$ is unique. 
\end{proof}

Now we will use these lemmata to derive the pushthrough condition
\begin{equation}
\label{eq:general_push_through_SM}
    U_j \cdot A \doteq \, v_{j-1}^\dagger\, A\, v_j,
\end{equation}

To do this, note that the $L$-periodic structure of the symmetry unitary implies that the unitary defined on $kL$ sites acts as an `ordinary' global uniform symmetry on an MPS defined on $k$ sites, with "lumped" MPS tensors $\mathcal{A}^{(\mathcal{S})}=A^{s_1}\cdots A^{s_L}$ (where we define the multi-index $\mathcal{S}=(s_1,\ldots,s_L)$). We may therefore apply the standard MPS push-through theorem to conclude that
\begin{equation}\label{eq:global_pushthrough}
\mathcal{U}_{L}\cdot\mathcal{A}=e^{i\phi}\mathcal{V}\mathcal{A}\mathcal{V}^\dag.
\end{equation}
The same condition holds after blocking any $p$ consecutive $L$-site unit cells, which we denote the tensor by $\mathcal{A}^p$. The corresponding ordinary transfer matrix of $\mathcal{A}^p$ is then approximately given by

\ie\label{eq:fixed_point_transfer_matrix}
\vcenter{\hbox{\includegraphics[scale=0.25]{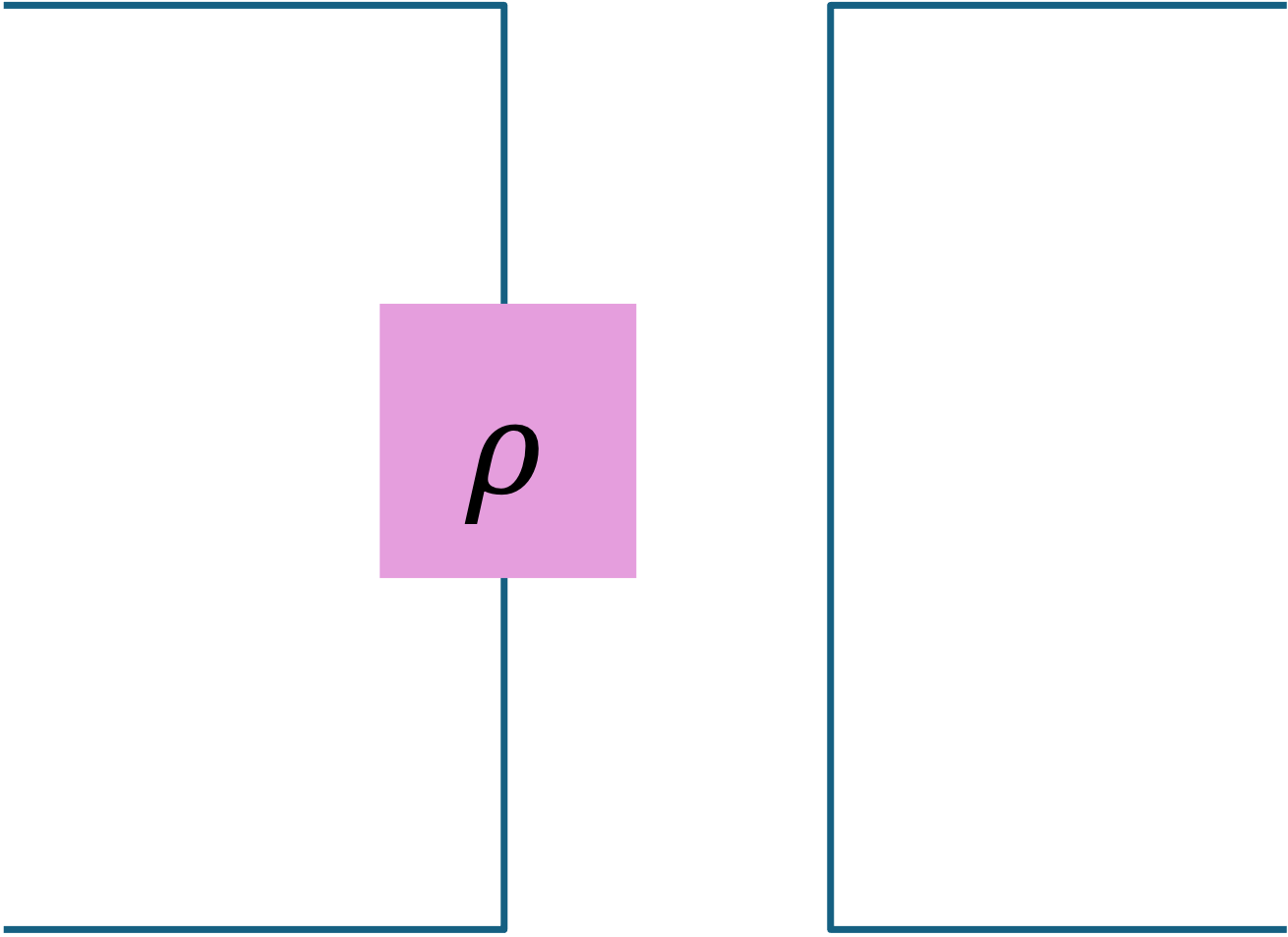}}}
\fe
with the error decaying as $O(e^{-pL/\xi})$. 
Now consider the total of $U$-transfer matrices for $\mathcal{A}^p$, which is given by $\mathbb{E}_{tot}=(\mathbb{E}_{U_1}\cdots\mathbb{E}_{U_L})^{\otimes p}$. By Eq.~\eqref{eq:global_pushthrough} and Eq.~\eqref{eq:fixed_point_transfer_matrix}, we have

\ie\label{eq:unit_inner_product}
\langle \mathcal{E}_{tot}(\mathcal{V}^\dag) , \mathcal{V}^\dag\rangle_T =  1,
\fe
where the inner product was defined in Eq.~\eqref{eq:generalized_inner_product}. Here Eq.~\eqref{eq:global_pushthrough} is true up to error of order $O(e^{- pL/\xi})$, and we can take $p$ to be sufficiently large (without changing $L$ which preserves the modulated symmetry structure) such that this error is negligible. Now we can use the Cauchy-Schwarz inequality to show:

\begin{equation}\label{eq:final_cauchy_schwarz_send}
\implies 1  = \abs{\langle \mathcal{E}_{tot}(\mathcal{V}^\dag) , \mathcal{V}^\dag\rangle_T}^2 \leq \abs{\langle \mathcal{V}^\dag , \mathcal{V}^\dag\rangle_T}^2 \mathnorm{\mathbb{E}_{tot}}^2 =\mathnorm{\mathbb{E}_{tot}}^2 \leq 1.
\end{equation}

Thus $\mathnorm{\mathbb{E}_{tot}}=1$, and by Lemma \ref{lem:3} there exist vectors $v_j$ such that $v^\dag_{j-1} \mathbb{E}_{U_{(j)}} \dot{=} v^\dag_{j}$, with $v^\dag_0 = \mathcal{V}^\dag$. Additionally, by the Cauchy-Schwarz equality, we have $\mathcal{E}_{tot}(\mathcal{V}^\dag) = \mathcal{V}^\dag$ so $v^\dag_0 \dot{=} v^\dag_L$. But since $\mathcal{E}(v^\dag_{j-1}v_{j-1}) =v^\dag_{j}v_{j}$, and $v^\dag_{0}v_{0}=\mathbbm{1}$, we automatically get that all the $v$'s are unitary. 

Then for each $\mathbb{E}_{U_{j}}$, Lemma \ref{lem:2} applies with the "weak push-through" condition Eq.~\eqref{eq:weak_pushthrough} becoming the strong push-through condition that is the statement we sought out to prove. Furthermore the periodicity requirement $v^\dag_0 \dot{=} v^\dag_L$ also follows. $\Box$ 

\subsection{Open Boundary Generalizations}

So far, we have provided an alternative proof for the result in Ref.~\cite{stephen2019subsystem}, applicable for modulated symmetries whose unitaries repeat after $L$ sites (these are known in the literature as $L$-cycle symmetries). This assumption on the structure of the MPS and the unitary is what gives rise to Eq.~\eqref{eq:global_pushthrough}. Our proof strategy, however, leads to a generalization to symmetries of \emph{open boundary} MPS that do not fall within the purview of $L$-cycle symmetries. \hfill \break

To set the stage, let $A^{s}$ be a left-canonical injective MPS tensor and let $\mathcal{U}=\bigotimes_{j=1}^L U_j$ be an on-site unitary that preserves the following $\chi^2$ open-chain states\footnote{One can verify that as $L\to\infty$, these are orthonormal states.} 

\begin{equation}\label{eq:OBC_MPS_BASIS}
    \ket{\psi_{\alpha\beta}} = \frac{1}{\sqrt{\lambda_\alpha}}\sum_{s_1\ldots s_L}\bbra{\alpha}A^{s_1}\cdots A^{s_L}\bket{\beta}\ket{s_1\ldots s_L},
\end{equation}
where $\bbra{\alpha}$ and $\bket{\beta}$ are standard basis vectors in bond space which diagonalize the Schmidt matrix; $\sqrt{\lambda_\alpha}$ is the corresponding Schmidt eigenvalue. This means that the matrix elements $\mathcal{U}_{\alpha'\beta';\alpha\beta} = \bra{\psi_{\alpha'\beta'}}\mathcal{U}\ket{\psi_{\alpha\beta}}$ should themselves define a $\chi^2\times \chi^2$ unitary matrix. We can then specify one of two conditions:

\begin{enumerate}
\item $\mathcal{U}$ satisfies the following pushthrough relation on the whole chain:
\begin{equation}\label{eq:slanted_pushthrough}
\mathcal{U}_{L}\cdot \mathcal{A} =e^{i\phi}\mathcal{V}\mathcal{A}\mathcal{W}^\dag
\end{equation}
with $\mathcal{V},\mathcal{W}$ unitaries in bond space and $\mathcal{A}$ being the $L$-site tensor produced by lumping $L$ copies of the original MPS tensor $A^{(s)}$. \hfill \break

\item The product of $U$-transfer matrices satisfies 
\begin{equation}\label{eq:global_op_norm}
   \mathnorm{\mathbb{E}_{U_1}\cdots \mathbb{E}_{U_L}} = \mathnorm{\mathbb{E}_{tot}}=1
\end{equation}
where the operator norm is induced by the inner product in Eq.~\eqref{eq:generalized_inner_product}.
\end{enumerate}

\begin{claim*}
    The above two conditions Eqs.~\eqref{eq:slanted_pushthrough} and \eqref{eq:global_op_norm} are equivalent as $L\to\infty$, and as a result each on-site unitary pushes through the MPS tensor as in Eq.~\eqref{eq:general_push_through_SM}.
\end{claim*}

\begin{proof}[\emph{\textbf{Proof:}}]
First, we show that the first condition Eq.~\eqref{eq:slanted_pushthrough} implies the second condition Eq.~\eqref{eq:global_op_norm} and also the pushthrough result Eq.~\eqref{eq:general_push_through_SM}. Following previous arguments, we can show that $L\to\infty$ implies $\langle \mathcal{E}_{tot}(\mathcal{V}^\dag) , \mathcal{W}^\dag\rangle_T = 1$ and by the Cauchy-Schwarz inequality, we get $\mathnorm{\mathbb{E}_{tot}}=1$. Thus Eq.~\eqref{eq:general_push_through_SM} holds with $v_0 = \mathcal{V}$ and $v_L = \mathcal{W}$. In this case, $\mathcal{U}_L$ may not define a symmetry on a periodic MPS if $\mathcal{V}\neq \mathcal{W}$, but with a suitable choice of $\bbra{\alpha},\bket{\beta}$, $\mathcal{U}_L$ will be a symmetry of $\ket{\psi_{\alpha\beta}}$. We also note that the implication of Eq.~\eqref{eq:general_push_through_SM} from Eq.~\eqref{eq:slanted_pushthrough} follows from results in Ref.~\cite{sahay2025classifying}, though our proof employs a different strategy.
\end{proof}

Now, we can show that Eq.~\eqref{eq:global_op_norm} implies Eq.~\eqref{eq:slanted_pushthrough}, thus Eq.~\eqref{eq:general_push_through_SM}. This is to say, we can prove Eq.~\eqref{eq:general_push_through_SM} without directly assuming any global pushthrough condition as in Eq.~\eqref{eq:global_pushthrough} or \eqref{eq:slanted_pushthrough}, but instead by assuming a condition on the properties of $U$-transfer matrices. To do this, we need to prove some additional lemmata.

\begin{lemma} \label{lem:4}
Recall from Lemma \ref{lem:3} that we have each $\mathbb{E}_{U_j}$ has $v^\dag_j,w^\dag_j$ satisfying Eq.~\eqref{eq:Generalized_Transfer_Condition} with $\abs{\lambda}=1$ , $w_{n-1}^\dag = v_n^\dag$. If the MPS is injective, then we have $w_n^\dag w_n = \mathbbm{1} + O(De^{-n/\xi})$ where $\xi$ is the correlation length of the MPS and $D$ is a constant that depends only on $w_0 = v$ -- hence, it depends only on the first on-site unitary $U_1$. Furthermore, there exists a finite $n_0$ depending on $D$ and $\xi$  after which $w_n$ is invertible for $n> n_0$. The same bounds apply for $v_n^\dag v_n$.
\end{lemma}

\begin{proof}[\emph{\textbf{Proof:}}] Simple corollary of Lemma \ref{lem:3}, since $w_n^\dag w_n = \mathcal{E}(w_{n-1}^\dag w_{n-1})$. Due to the condition $\langle w_0^\dag, w_0^\dag\rangle_T = 1$, we have $w_0^\dag w_0 = \mathbbm{1} + \mathcal{S}$ where $\mathcal{S}$ are sums of subdominant eigenvectors/Jordan chains of the transfer matrix which all satisfy $\Tr(\rho \mathcal{S})=0$. $\mathbbm{1}$ is the unique fixed point of the channel $\mathcal{E}(\cdot)$ which is thus reached through iteration exponentially quickly.
\end{proof}
\begin{lemma}\label{lem:5}
If a product of $U$-transfer matrices $\mathbb{E}_{tot} = \mathbb{E}_{U_1}\dots \mathbb{E}_{U_n}$ has $\mathnorm{\mathbb{E}_{tot}}=1$, then there exists a rank-$1$ matrix $\bket{R}\bbra{L}$ in bond space such that

\begin{equation}\label{eq:rank_1_approx}
    \mathnorm{ \mathbb{E}_{tot} -  \bket{R}\bbra{L}} \leq O(D'ne^{-n/2\xi})
\end{equation}
where the finite constant $D'$ depends only on properties of the original transfer matrix $\mathbb{E}$ and the left-most unitary $U_1$. The operator norm we use doesn't matter, as all norms on $\mathbb{C}^n$ are equivalent.
\end{lemma}

\begin{proof}[\emph{\textbf{Proof:}}] To start, let us review the proof of the corresponding result for the original transfer matrix $\mathbb{E}$ (i.e when $U=\mathbbm{1}$). We can take any ket $\bket{\beta}$ and bra $\bbra{\alpha}$ in bond space and decompose them into the left- and right-eigenvectors (or, in the case of Jordan blocks, properly normalized bases for the eigenspaces) respectively, which form complete bases for bond space and are dual-orthonormalized. It follows that when $\alpha\neq \mathbbm{1}$ and $\beta \neq \rho$, the matrix element $\abs{\bbra{\alpha}\mathbb{E}^n\bket{\beta}} \leq Ce^{-(n-n_J)/\xi}$ where $\xi^{-1} = -\ln(\lambda_2)$ with $\lambda_2$ being the largest eigenvalue less than $1$. The number $n_J$ is the size of the largest Jordan block and $C$ also depends on this largest Jordan block -- thus $n_J$ and $C$ only depend on the bond dimension $\chi$ since $n_J$ is upper bounded by $\chi-1$. Choosing a matrix norm computed by summing up norms of the matrix elements, we have

\begin{equation}
\mathnorm{\mathbb{E}^n - \bket{\rho}\bbra{\mathbbm{1}}} \leq Ce^{-n/\xi}
\end{equation}
where $C$ depends on the bond dimension $\chi$ and the correlation length $\xi$. \hfill \break
\end{proof}
Now we are ready to tackle the case of $\mathbb{E}_{tot}$. By Lemma \ref{lem:4}, each $v_j,w_j$ is unitary up to some error $O(Ce^{-j/\xi})$. Accordingly, we approximately recover the push-through relation

\begin{equation}\label{eq:approx_pushthrough}
    U_jA = v_{j}Av_{j+1}^\dag + O(De^{-j/\xi})
\end{equation}
where we used the notation of Lemma \ref{lem:3} to write $ v_j = w_{j-1}$. Here, we will take the $v$'s to be \textbf{exactly} unitary, and simply absorb the $O(De^{-j/\xi})$ deviation from unitarity into the overall error term. Now consider the latter half of the chain of $U$-transfer matrices, i.e. $\mathbb{E}_{U_{n/2}}\dots \mathbb{E}_{U_n}$. By Eq.~\eqref{eq:approx_pushthrough} and $v_j^\dag v_j =\mathbbm{1}$, we have

\begin{equation}
   \mathbb{E}_{U_{n/2}}\dots \mathbb{E}_{U_n} = v_{n/2}\mathbb{E}^{n/2}v_{n+1}^\dag + O(Dne^{-n/2\xi}).
\end{equation}

By previous logic, we have 

\begin{align}
    v_{n/2}\mathbb{E}^{n/2}v_{n+1}^\dag &= w_{n/2-1}\bket{\rho}\bbra{\mathbbm{1}} v_{n+1}^\dag + O(Ce^{-n/2\xi}) + O(Dne^{-n/2\xi}) \\ 
    &= v_{n/2}\bket{\rho}\bbra{\mathbbm{1}} v_{n+1}^\dag + O(D'ne^{-n/2\xi}).
\end{align}

The first term is a rank-1 matrix. Recall that any product involving a rank-1 matrix is also rank-1. This implies that the full product is

\begin{equation}
    \mathbb{E}_{U_{1}}\dots \mathbb{E}_{U_n} = \bket{R}\bbra{v_{n+1}^\dag}  + O(D'ne^{-n/2\xi})
\end{equation}
where we used the fact that the operator norm of any product of $\mathbb{E}_U$'s is bounded from above by a universal constant (depending on our choice of norm). The lemma follows. \hfill \break

Finally, we can prove the claim that Eq.~\eqref{eq:global_op_norm} implies Eq.~\eqref{eq:slanted_pushthrough}. We can compute the matrix elements of $\mathcal{U}$ within the subspace of open MPS defined by Eq.~\eqref{eq:OBC_MPS_BASIS}. We have
\begin{equation}\label{eq:Rank-1-transfer-capstone}
\mathcal{U}_{\alpha'\beta';\alpha\beta} = \bra{\psi_{\alpha'\beta'}}\mathcal{U}\ket{\psi_{\alpha\beta}} = \bbra{\alpha,\alpha'}\bket{R}\bbra{L}\bket{\beta,\beta'} = R_{\alpha\alpha'}L_{\beta\beta'}
\end{equation}
by Lemma \ref{lem:5} and the definition of the $U$-transfer matrix. Notice that the matrix element $R_{\alpha\alpha'}$ must be a unitary matrix $\mathcal{V}$ as well as $L_{\beta\beta'}$ (corresponding to $\mathcal{W}$). These unitaries can be taken to act on the left and right bonds of the product of $L$ MPS tensors. Thus, we recover Eq.~\eqref{eq:slanted_pushthrough} and the claim follows.

\section{Examples of Symmetry Push-through for Various Modulated Symmetries}
\label{app:examples-for-push-throughs}

\subsection{General Modulated Symmetries}

Now, we will describe the framework for deriving push-through rules for general modulated symmetries. We use the framework in Refs.~\cite{pace2026spacetime,pace_bulmash_2026_LSM_Modulated,bulmash2025defect}, where modulated symmetries are defined by a semidirect product $G\rtimes \mathcal{C}$, where $G$ is an "on-site" symmetry whose group operators are realized as tensor products of single-site unitaries, and $\mathcal{C}$ is a crystalline symmetry group. Here, we always fix $\mathcal{C}=\mathbb{Z}_L$ (this special case corresponds to $L$-cycle symmetries as stated in Refs.~\cite{sauerwein2019matrix,stephen2019subsystem}) and thus define the semidirect product by mapping the single-site translation $\mathcal{T}\in \mathbb{Z}_L$ to an automorphism $\mathcal{T}(\cdot) \in \mathrm{Aut}(G)$. Then for $g\in G$, on-site unitaries obey 

\begin{equation}\label{eq:nonabelian_group_automorphism}
\mathcal{T}\mathcal{U}_g\mathcal{T}^\dag = \mathcal{U}_{\mathcal{T}(g)}.
\end{equation}
Since we require $\mathcal{T}^L(\cdot)$ to be the identity automorphism, the unitaries $U_{g,j}$ in $\mathcal{U}$ must be $L$-periodic and the push-through theorem holds; furthermore Eq.~\eqref{eq:nonabelian_group_automorphism} implies that we can write 

\begin{equation}\label{eq:on_site_unitary_fixed_lcycle}
\mathcal{U}_g=\bigotimes_{j=1}^LU_{g,j} = \bigotimes_{j=1}^LU_{\mathcal{T}^{j-1}(g)}.
\end{equation}
 Now we consider the \textit{restricted} action of the unitary $\mathcal{U}_{(gh)^{-1}}\mathcal{U}_{h}\mathcal{U}_{g}$ on $\ell \geq \ell_0$ sites of the MPS, where $\ell_0$ is the injective length \footnote{The 'injective length' of an MPS is the minimal $\ell_0$ such that the product of MPS tensors $A_{\alpha_0,\alpha_1}^{(j_1)}A_{\alpha_1,\alpha_2}^{(j_2)}\ldots A_{\alpha_{\ell_0-1},\alpha_{\ell_0}}^{(j_{\ell_0})}  = M_{\alpha_0,\alpha_{\ell_0}}^{(j_1\ldots j_{\ell_0})}$ spans the entire vector space of matrices in bond space (upon choosing specific indices $j_i$ and taking linear combinations). For an injective MPS, such an $\ell_0<\infty$ always exists and such a condition is satisfied for any $\ell\geq \ell_0$ \cite{mps_representations}.}. For now, assume that the on-site unitaries are all linear representations of $G$. Then this action is trivial; however, if we use the push-through condition, we get

\begin{equation}
    A^{s_j}\cdots A^{s_{j'}} = v_{j}^\dag((gh)^{-1})v_{j}^\dag(h)v_{j}^\dag(g)A^{s_j}\cdots A^{s_{j'}}v_{j'}(g)v_{j'}(h)v_{j'}((gh)^{-1}) \prod_{l=j+1}^{j'}e^{i\mathrm{d}\phi_l(g,h)}
\end{equation}
with $j'-j\geq \ell_0$, and $\mathrm{d}\phi_l(g,h) = \phi_l(g)+\phi_l(h)-\phi_l(gh)$ being a coboundary. We can take linear combinations and contract physical legs such that we can replace $A\cdots A$ with any bond space matrix we like. Choose $A\cdots A \to \mathbbm{1}$ and thus conclude that 

\begin{equation}\label{eq:general_cocycle_constraint}
v_{j'}(g)v_{j'}(h)v_{j'}(h^{-1}g^{-1})\prod_{l=j+1}^{j'}e^{i\mathrm{d}\phi_l(g,h)} = v_{j}(g)v_{j}(h)v_{j}(h^{-1}g^{-1}) = \mathcal{V}.
\end{equation}
Now choose $A\cdots A \to M$ for arbitrary bond matrix $M$; Schur's lemma thus implies that $\mathcal{V} = e^{i\phi} \mathbbm{1}$. This implies that 

\begin{equation}
    \begin{split}
        v_{j}(g)v_{j}(h) = \omega_j(g,h) v_{j}(gh),\quad
        \omega_{j'}(g,h)\omega_j^{-1}(g,h) = \prod_{l=j+1}^{j'}e^{i\mathrm{d}\phi_l(g,h)},
    \end{split}
\end{equation}
where we have redefined the phase and made the dependence on the group elements explicit. The notation is chosen precisely because the bond unitaries $v_j(g)$ form a possibly projective representation of $G$, and $\omega(g,h)$ represents the cocycle in $H^2(G,U(1))$. Notably, the element in group cohomology does not depend on the unit cell, since $\omega_j(g,h)$ only changes by coboundaries as $j$ varies. \hfill \break

Now we can compare the action of $\mathcal{U}_{\mathcal{T}(g)}$ to that of $\mathcal{T}\mathcal{U}_{g}\mathcal{T}^{-1}$. This leads to the equality

\begin{equation}\label{eq:compare_g_tg}
    v_j^\dag(\mathcal{T}(g))A\cdots Av_{j'}(\mathcal{T}(g)) \prod_{l=j+1}^{j'}e^{i\phi_l(\mathcal{T}(g))} = v_{j+1}^\dag(g)A\cdots Av_{j'+1}(g) \prod_{l=j+2}^{j'+1}e^{i\phi_l(g)}.
\end{equation}

Following the previous injectivity argument and using translation invariance of the cocycle $\omega(g,h)$, we have
\begin{equation}\label{eq:Grand_Pullback_Cocycle_Constraint_stage_1}
\begin{split}
v_{j+1}({g}) = e^{i\varphi_j({g})}v_{j}(\mathcal{T}({g})), \quad 
 e^{i\varphi_{j'}({g})}e^{-i\varphi_j({g})} =  \prod_{l=j+1}^{j'}e^{i(\phi_l(\mathcal{T}(g)) -\phi_{l+1}(g))}.
\end{split}
\end{equation}

Now recognize that because of the constraint Eq.~\eqref{eq:on_site_unitary_fixed_lcycle}, the pushthrough phases $\phi_j(g)$ satisfy $\phi_j(g) = \phi_{j-1}(\mathcal{T}(g))\implies \phi_j(g) = \phi_0(\mathcal{T}^j(g))$. Thus, it turns out that $\varphi_j(g)$ is independent of $j$, and we have

\begin{equation}\label{eq:Grand_Pullback_Bond_Constraint}
\begin{split}
v_{j+1}({g}) &= e^{i\varphi({g})}v_{j}(\mathcal{T}({g})).
\end{split}
\end{equation}

As a consequence, we have
\begin{equation}\label{eq:Grand_Pullback_Cocycle_Constraint}
\begin{split}
[\omega(g,h)] = [\omega(\mathcal{T}({g}),\mathcal{T}({h}))]
\end{split}
\end{equation}
where the equality of $H^2(G,U(1))$ classes $[\omega]$ in the second line is up to $G$ coboundaries $\varphi({g})+\varphi({h}) - \varphi({g}{h})$. The first line of Eq.~\eqref{eq:Grand_Pullback_Cocycle_Constraint} provides a recipe for solving for the specific bond unitaries $v_j({g})$ for each modulated symmetry group. The second line yields the same constraint on strong modulated SPTs that is found using the SymTFT framework \cite{pace2026spacetime} or real-space defect network constructions \cite{bulmash2025defect}. \hfill \break

\subsection{Classification of Weak SPTs}
In Eqs. \eqref{eq:Grand_Pushthrough_App} and \eqref{eq:Grand_Pullback_Bond_Constraint}, there are two phases $\varphi(g)$ and $\phi_j(g) = \phi(\mathcal{T}^j(g))$. First we note that by redefining $v_j(g)$, we can absorb $\varphi(g)$ into $\phi_j(g)$ in a way that preserves $\phi_j(g) = \phi_{j-1}(\mathcal{T}(g))$. For ordinary unmodulated symmetries, $e^{i\phi_j(g)}$  is site-independent and  
forms a linear representation of the on-site group $G$ (i.e. an element of $H^1(G,U(1))$). 
This can be interpreted as decorating a charge of $G$ on each site, which yields the classification of weak SPTs in 1+1 dimensions. 
\hfill \break

For modulated symmetries, the pushthrough phases vary along the chain as $e^{i\phi_j(g)} = e^{i\phi_0(\mathcal{T}^j(g))}$. Translation shifts the phase $e^{i\phi(g)}\to e^{i\phi(\mathcal{T}(g))}$, hence the assignment of the $H^1(G,U(1))$ on a unit cell transforms accordingly. Since translation should not change the weak SPT phase due to MPS being translation invariant, we need to identify elements $e^{i\phi(g)},e^{i\phi(\mathcal{T}(g))}\in H^1(G,U(1))$. Thus the classification of weak SPTs is given by the quotient group

\begin{equation}\label{eq:weak_SPT_classification}
H^1(G,U(1))/\{e^{i\phi(g)}\sim e^{i\phi(\mathcal{T}(g))}\}.
\end{equation}
Along with the classification of strong SPTs in Eq.~\eqref{eq:Grand_Pullback_Cocycle_Constraint}, this matches the result in Ref.~\cite{bulmash2025defect}, and thus verifies the crystalline equivalence principle for modulated symmetries within the MPS framework.

\subsection{Dipole Symmetry}

Besides the exponential symmetry discussed in the main text, another important example of a modulated symmetry is a dipole symmetry $G$, given by

\begin{equation}\label{eq:Dipole-Sym}
    \mathcal{U}_{D,g} = \bigotimes_{j=1}^L U^{j}_{g}, \quad \mathcal{U}_{C,g} = \bigotimes_{j=1}^L U_{g}.
\end{equation}

We show that the first unitary $U$ pushes through the MPS tensor in two different ways, depending on whether we invoke the push-through rule \eqref{eq:general_push_through_SM} for the dipole symmetry or the charge symmetry. The push-through conditions give 

\begin{equation}
    U_g \cdot A \doteq v_c^\dag(g) A v_c(g), \quad U^j_g \cdot A \doteq v^\dag_{d,j-1}(g) A v_{d,j}(g),
\end{equation}
which implies

\begin{equation}
U_g\cdot A  \doteq v_c^\dag(g) A v_c(g)  \doteq v_{d,j}^\dag(g) v_{d,j-1}(g) A v_{d,j}^\dag(g) v_{d,j+1}(g),
\end{equation}
where the projective representations of the (unmodulated) charge and (modulated) dipole symmetries are labeled by $v_c(g)$ and $v_{d,j}(g)$ respectively. Note that the former is site-independent and hence the subscript $j$ is omitted.

By the same argument as before, we can obtain that 
\begin{equation}v_{d,j}^\dag(g)  \doteq v^\dag_c(g) v_{d,j-1}^\dag(g)\doteq (v^\dag_c(g))^j v_{d,0}^\dag(g),
\end{equation} 
as depicted in Fig. \ref{fig:dipole_pushthrough}. 
Now observe that we have

\ie
U(g)\cdot A \doteq v^\dag_c(g) A v_c(g) &\doteq v^\dag_{d,0}(g) A v_{d,1}(g)= v^\dag_{d,0}(g) A v_{d,0}(g)v_c(g) \\
    &\implies v_c^\dag(g) A \doteq v^\dag_{d,0}(g) A v_{d,0}(g).\label{eq:Horizontal_Pushthrough_Dipole}
\fe

This is, up to conventions, the same push-through rule derived in Ref.~\cite{ho_tat_lam_dipolar_SPT_classification}.

\begin{figure}[h]
\centering
\begin{tikzpicture}[
  x=0.85cm,y=0.85cm,
  tensor/.style={draw, rounded corners=1.5pt, minimum size=7.5mm, thick, fill=gray!10},
  leg/.style={thick},
  op/.style={draw, circle, inner sep=0.6pt, thick, fill=white},
  every node/.style={font=\scriptsize},
]

%-------------------------
% Panel (b): charge (same y-level as panel a)
%-------------------------
% Horizontal offset between panels (tune if your column is very narrow)
\coordinate (M) at (3.8,0);
 
\node[tensor] (B1) at ($(M)+(0,0)$) {$A$};
\draw[leg] (B1.north) -- +(0,0.4) node[op,above] {$\ \ U\ \ $} -- +(0.,1.6);
\draw[leg] (B1.west)  -- +(-0.85,0);
\draw[leg] (B1.east)  -- +(0.85,0);
 
\node[font=\small] (eq2) at ($(M)+(1.8,0)$) {$\doteq$};

\node[tensor] (B2) at ($(M)+(4.8,0)$) {$A$};
\draw[leg] (B2.north) -- +(0,1.4);
\draw[leg] (B2.west)  -- +(-0.4,0) node[op,left] {$\ \ v^\dag_c\ \ $}  -- +(-2,0);
\draw[leg] (B2.east)  -- +(0.4,0)  node[op,right] {$\ \ v_c\ \ $} -- +(+2,0);

%-------------------------
% Panel (c): dipole (same y-level as panel a)
%-------------------------
% Horizontal offset between panels (tune if your column is very narrow)
\coordinate (M) at (9.8,0);

\node[font=\small] (eq2) at ($(M)+(1.8,0)$) {$\doteq$};
 
\node[tensor] (B2) at ($(M)+(4.8,0)$) {$A$};
\draw[leg] (B2.north) -- +(0,1.4);
\draw[leg] (B2.west)  -- +(-0.4,0) node[op,left] {$(v^\dag_c)^j v^\dag_{d,0}$}  -- +(-2,0);
\draw[leg] (B2.east)  -- +(0.4,0)  node[op,right] {$v_{d,0} v_c^{j+1}$} -- +(+2,0);

\end{tikzpicture}
\vspace{-0.1in}
\caption{ Push-through rule for dipole symmetries. }
\label{fig:dipole_pushthrough}
\end{figure}
To streamline the result, we can reformulate the dipole symmetry by treating it as an abelian on-site symmetry group $G\times G$ and the automorphism $\mathcal{T}(g_c,g_d) = (g_cg_d,g_d)$. This yields $v_{j+1}(g_c,g_d) \doteq v_{j}(g_cg_d,g_d)$ (note that $v_j$ now labels projective representations of $G\times G$). At the cost of phases which always cancel out on pairs of bonds in Eq.~\eqref{eq:general_push_through_SM}, we can write $v_{j}(g_c,g_d) = v_j(g_c,\mathbbm{1})v_j(\mathbbm{1},g_d) = v_{d,j}(g_d)v_c(g_c)$ where $\mathbbm{1}$ is the identity element. Then we have $v_{d,j}(g_d)\doteq v_{d,0}(g_d)v^j_c(g_d)$. This reproduces Fig.~\ref{fig:dipole_pushthrough}, in particular Eq.~\eqref{eq:Horizontal_Pushthrough_Dipole}, from which the classification of dipolar SPTs follows (Eq. 1.1 in \cite{ho_tat_lam_dipolar_SPT_classification}). 

A generalization to multipolar symmetries is also clear, following \cite{Saito_Multipolar_MPS}. Here, we consider a multipolar symmetry group given by on-site group $G^p$ with group homomorphism $\mathcal{T}(g_1,g_2,\ldots,g_p) = (g_1g_2,g_2g_3,\ldots,g_p)$. For example, $g_1,g_2,g_3$ represents the elements of global charge symmetry, the dipole symmetry and the quadrupole symmetry, respectively. Following the previous strategy to solve for the pushthrough tensors $v_{1,j}(g_1)v_{2,j}(g_2)\cdots v_{p,j}(g_j)$ in terms of the group elements, we recover the result 
\begin{equation}
v_{n,0}^\dag(g) A \doteq v_{n+1,0}^\dag(g) A v_{n+1,0}(g),\quad 1\le n\le p-1 ,
\end{equation}
which was derived in Ref.~\cite{Saito_Multipolar_MPS}.

\subsection{Tilted Exponential Symmetry}
Inspired by the case of MPS with both charge and dipole symmetry, we suggest a symmetry group that consists of exponential symmetry and "dipole-exponential" symmetry \cite{yan2024generalized}:

\begin{align}\label{eq:Tilted_exponential_unitaries}
    \mathcal{U}_{E,g} &= \bigotimes_j U^{b^{j-1}}_g, \\
    \mathcal{U}_{DE,g} &= \bigotimes_j U^{jb^{j-1}}_g.
\end{align}

One can readily verify that 
$\mathcal{T}\mathcal{U}_{E,g}\mathcal{T}^{-1} = \mathcal{U}_{E,g}^b$ and 
\begin{equation}\label{eq:tilted_exp_algebra}
\mathcal{T}\mathcal{U}_{DE,g}\mathcal{T}^{-1} = \bigotimes_j U^{(j+1)b^{j}}_g = \bigotimes_j (U^{b^{j-1}}_g)^b (U^{jb^{j-1}}_g)^b = \mathcal{U}_{E,g}^b\mathcal{U}_{DE,g}^b   . 
\end{equation}

Demanding that $\mathcal{U}_{E,g}$ is well-defined on a periodic chain of length $L$ leads to $U^{b^L-1}_g = \mathbb{I}$. Imposing the same on $\mathcal{U}_{DE,g}$, we obtain 

\begin{equation}
    \mathcal{T}^L\mathcal{U}_{DE,g}\mathcal{T}^{-L} = \bigotimes_j U^{(j+L)b^{j-1+L}}_g = \bigotimes_j (U^{b^{j-1}}_g)^{Lb^L} (U^{jb^{j-1}}_g)^{b^L} = \mathcal{U}_{E,g}^{Lb^L}\mathcal{U}_{DE,g}^{b^L} = \mathcal{U}_{DE,g}.
\end{equation}

Using $U^{b^L-1}_g = \mathbb{I}$, we have

\begin{equation} 
\mathcal{U}_{E,g}^{L} = \mathbb{I} \implies U^L_g = \mathbb{I}.
\end{equation}

Thus, if $G=\Z_N$ we conclude that $L=0~(\text{mod } N)$ and $b^L=1 ~(\text{mod } N)$ are necessary conditions for this symmetry to be realized on a periodic chain, which in turn imply $b$ must be coprime to $N$. In such a case, there is a family of solutions $L=k\varphi (N)N$ where $k\in\mathbb{N}$ and $\varphi (N)$ is the Euler’s totient function.

\hfill \break

Now we need to find bond unitaries $v_j$ corresponding to this symmetry action. %Let the element $(g_1,g_2)$ represents $\mathcal{U}_{E,g_1}\mathcal{U}_{DE,g_2}$.  %Let $\varrho: (g_1,g_2) \mapsto \mathcal{U}_{E,g_1}\mathcal{U}_{DE,g_2}$.
The push-through conditions give 
\begin{equation}
U^{b^{j-1}}_g \cdot A \doteq v_{e,j-1}^\dag(g) A v_{e,j}(g), \quad U^{jb^{j-1}}_g \cdot A \doteq v^\dag_{de,j-1}(g) A v_{de,j}(g).
\end{equation}
Due to the exponential symmetry  algebra, we  first have $v_{e,j}(g)\doteq v^{b^j}_{e,0}(g)$. Then by considering the push-through conditions of $U^{jb^{j-1}(b-1)}_g U^{b^{j}}_g= U^{-jb^{j-1}}_g U^{(j+1)b^j}_g$:
\begin{align}
\begin{split}
& U^{jb^{j-1}(b-1)}_g U^{b^{j}}_g\cdot A \doteq  v_{e,j}^\dag(g) (v^\dag_{de,j-1})^{b-1}(g) A v_{de,j}^{b-1}(g) v_{e,j+1}(g),\\ 
& U^{-jb^{j-1}}_g U^{(j+1)b^j}_g\cdot A\doteq v_{de,j}^{\dagger} (g) v_{de,j-1}(g) A v^\dag_{de,j}(g)v_{de,j+1} (g),
 \end{split}
\end{align}
we can obtain
\begin{equation}
v_{de,j}(g)\doteq  v^b_{de,j-1}(g)v_{e,j}(g)=v^{b}_{de,j-1}(g) v^{b}_{e,j-1}(g).
\end{equation}

If $G=\Z_N$, the projective representation can then be characterized by the coboundary-invariant phase 
$v_{e,j}v_{de,j}v^{\dagger}_{e,j}v^{\dagger}_{de,j}=\exp(2\pi ik /N)$ with $k\in\Z_N$ where $v_{e,j}$ and $v_{de,j}$ correspond to the bond unitaries associated with the generator of exponential symmetry and “dipole-exponential” symmetry. This  coboundary-invariant phase is site-independent as we have assumed the linear representation on each site. This results in the consistency condition $k(b^2-1)=0~ (\text{mod } N)$. Thus the corresponding SPT phases are classified by $\mathbb{Z}_{\gcd(b^2-1,N)}$. 

As a final note, we can recast this symmetry group as a semidirect product $(\mathbb{Z}_N\times\mathbb{Z}_N)\rtimes \mathbb{Z}_L$ with the homomorphism $\mathcal{T}(g_E,g_{DE}) =(g_E^bg_{DE}^b,g_{DE}^b)$ as dictated by Eq. \eqref{eq:tilted_exp_algebra}. Then the representation Eqs. \eqref{eq:Tilted_exponential_unitaries} is a faithful representation of this modulated symmetry group.
\subsection{Exponential symmetry with charge symmetry: Unfaithful Unitary Representations}

In the main text, the SPT and SPT-LSM classification results are derived solely from the algebraic structure of the modulated symmetry. In particular, the semidirect-product structure $G\rtimes \mathbb{Z}_L$ constrains the cocycles for on-site and bond unitaries. However, if the representation of the modulated symmetry group is unfaithful (but still on-site linear), additional constraints on the existence of SPTs can arise beyond those implied by the translation invariance of the cocycle. Here,we present an example with an unfaithful representation that exhibits such additional constraints.\hfill \break

To begin, consider the on-site group $G = G_1\times G_2=\mathbb{Z}_N\times\mathbb{Z}_N$ and define the modulated symmetry $G\rtimes\mathbb{Z}_L$ with the automorphism $\mathcal{T}(g_1,g_2) = \mathcal{T}(g_1^b,g_2)$ where $g_{1/2}\in G_{1/2}$. Suppose that the representation is linear on-site, so the constraint on the only nontrivial bond cocycles is $\omega(g_1,g_2)^{b-1}=1$. For simplicity, take $N=k(b-1)$, such that expect that strong SPTs are classified by $\mathbb{Z}_{b-1}$.\hfill \break

Now consider the following generating unitaries for a symmetry that realizes the above algebraic structure:

\begin{equation}\label{eq:Exp_Charge_2_Appendix}
\begin{split}
\mathcal{U}_E = \bigotimes_{j=1}^L U^{b^{j-1}},\quad 
\mathcal{U}_C = \bigotimes_{j=1}^L U.
\end{split}
\end{equation}
Here, the exponential and charge symmetries $\mathcal{U}_E$ and $\mathcal{U}_C$ are built from the \textit{same} unitary operator acting on physical sites. Again, let $U^{N}=\mathbbm{1}$ with $N=k(b-1)$. This implies a nontrivial relationship between the bond unitaries that they push through to:

\ie\label{eq:ec_correlated_1}
\vcenter{\hbox{\includegraphics[scale=0.3]{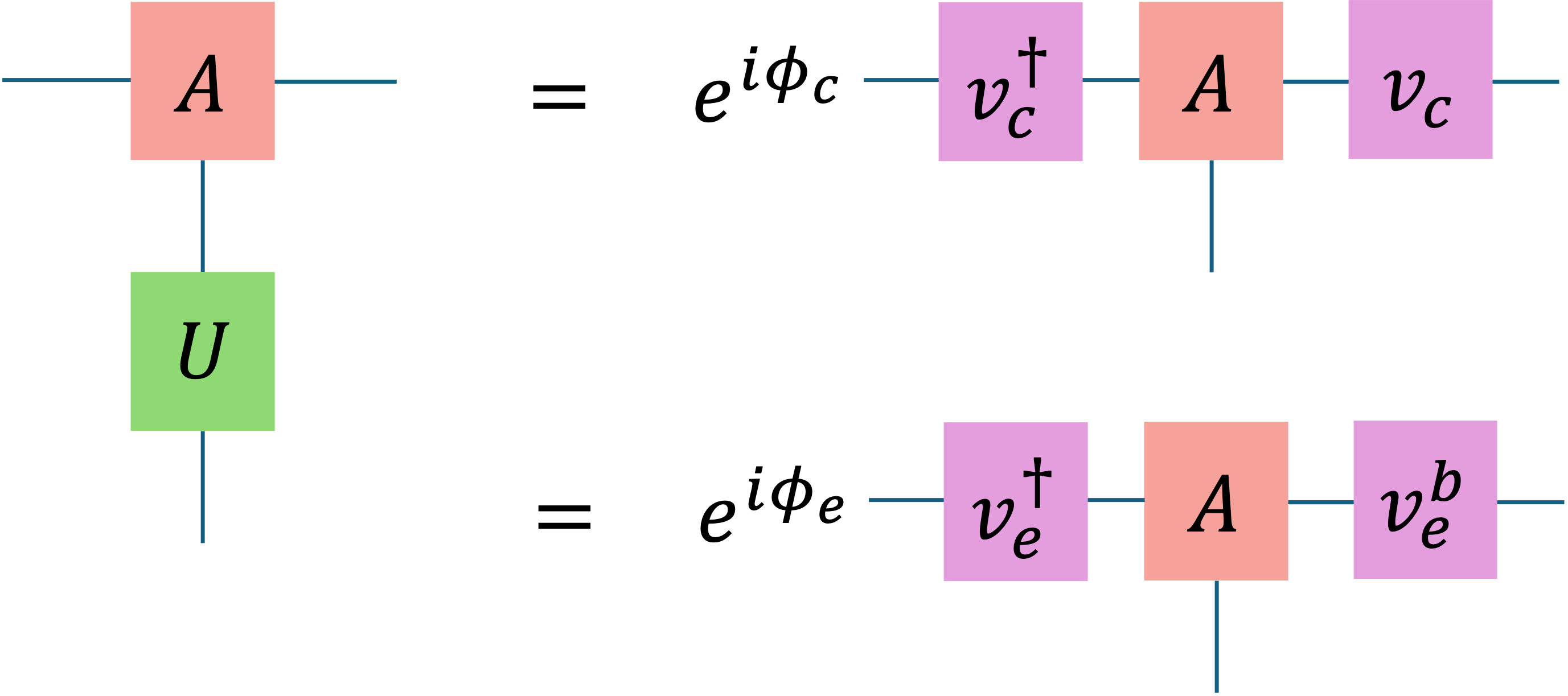}}}.
\fe
This implies that
\ie\label{eq:ec_correlated_2}
\vcenter{\hbox{\includegraphics[scale=0.3]{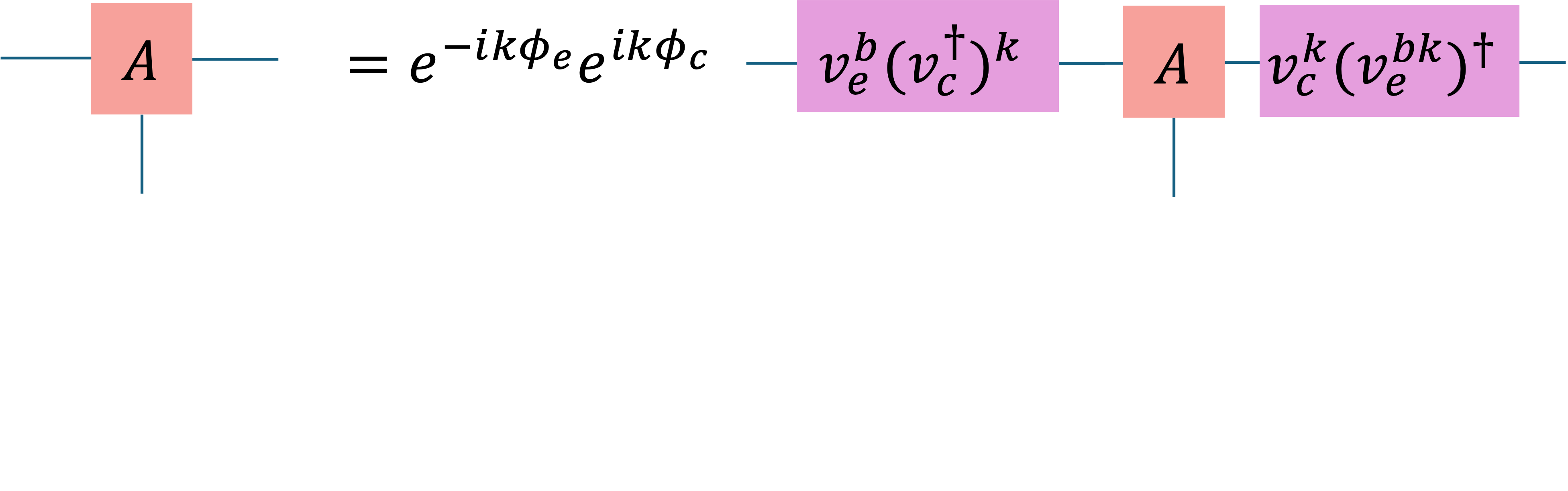}}},
\fe
which in turn lets us rewrite the left-normalization condition \eqref{eq:canonical} of the transfer matrix as 

\ie\label{eq:ec_rewritten_transfer}
\vcenter{\hbox{\includegraphics[scale=0.3]{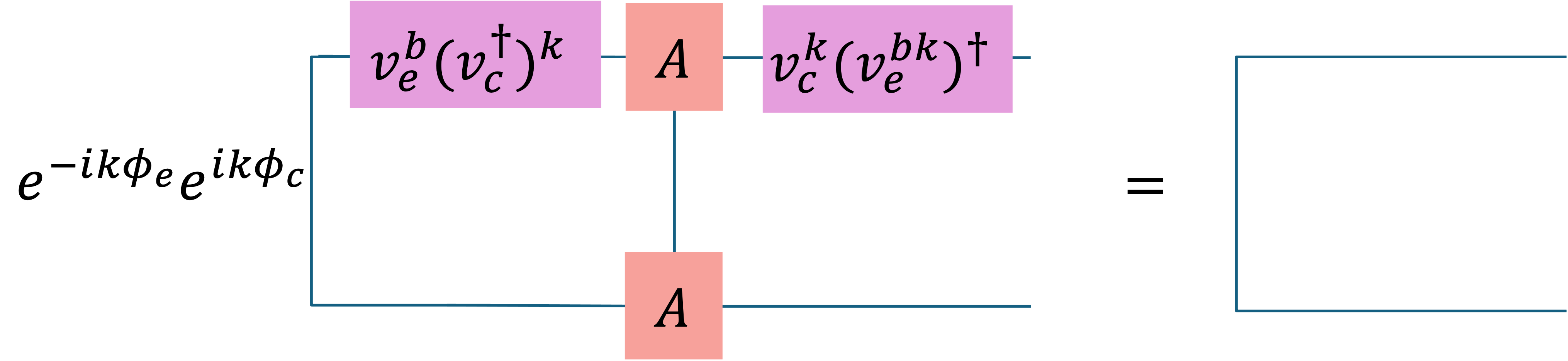}}}.
\fe
Rearranging terms, we have
\ie\label{eq:ec_rewritten_transfer}
\vcenter{\hbox{\includegraphics[scale=0.3]{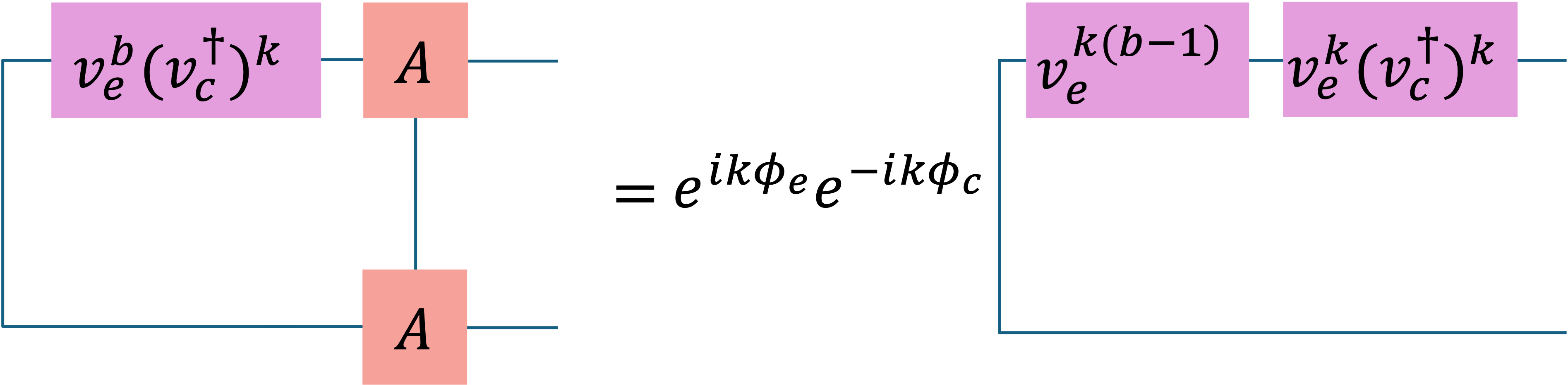}}}.
\fe
Now note that $v_e^{k(b-1)}=e^{i\varphi}$; one can show this by applying $\mathcal{U}_E^{k(b-1)} = \mathbbm{1}$ to a chain of MPS tensors longer than the injective length and applying the usual argument. As a consequence of injectivity, the dominant left eigenvector on the unit circle of the transfer matrix is unique (with eigenvalue equal to $1$) and thus $v_e^k= v_c^k$ with $e^{ik(\varphi+\phi_e-\phi_c)}=1$. \hfill \break

Notice that the condition $v_e^k=v_c^k$ constrains the possible projective phase between the $v_e$ and $v_c$. Indeed, if we have $v_ev_c=\omega_{ce}v_cv_e$ with $\omega(g_1,g_2)=\omega^{g_1g_2}_{ce}$, then consistency requires $\omega_{ce}^{k}=1$. This is an \emph{additional} constraint independent of $\omega_{ce}^{b-1}=1$ since the latter is derived from Eq.~\eqref{eq:Grand_Pullback_Cocycle_Constraint}. \hfill \break

This additional constraint is a consequence of the representation of the exponential-charge symmetry in Eq.~\eqref{eq:Exp_Charge_2_Appendix} being unfaithful. Indeed, the physical symmetry operators satisfy the relation $\mathcal{U}_E^k = \mathcal{U}_C^k$. 
Hence the representation map from the on-site group  $G = \mathbb{Z}_N\times\mathbb{Z}_N$ to physical unitaries is not injective.
This unfaithfulness is present even when the on-site unitaries themselves are a faithful representation of $\mathbb{Z}_N$. 
The faithfully represented subgroup of the modulated symmetry group $(\mathbb{Z}_N\times\mathbb{Z}_N)\rtimes \mathbb{Z}_L$ is instead
 $(\mathbb{Z}_N\times\mathbb{Z}_k)\rtimes \mathbb{Z}_L$ with the automorphism $\mathcal{T}(g_1,g_2) = (g_1g_2^{b-1},g_2^b)$. This automorphism yields the constraint $\omega(g_1,g_2)^{b-1}=1$. On the other hand, since $H^2(\mathbb{Z}_N\times\mathbb{Z}_k,U(1)) = \mathbb{Z}_{\mathrm{gcd}(N,k)} = \mathbb{Z}_{k}$, the cocycle must also satisfy $\omega(g_1,g_2)^{k}=1$. For generic $b,N$, one can show that the symmetry in Eq.~\eqref{eq:Exp_Charge_2_Appendix} realizes an on-site $\mathbb{Z}_N\times\mathbb{Z}_{N/\mathrm{gcd}(N,(b-1))}$ symmetry.

\section{LSM and SPT-LSM theorem}
\label{app: LSM}
%On-Site Projective Representations}
To discuss LSM-type obstructions, we suppose that the Hilbert space of each site carries a projective representation labeled by a cocycle $\nu(g,h)$. 
Due to translation action, $\mathcal{T} U_{g,j}\mathcal{T}^{\dagger}=U_{\mathcal{T}(g),j-1}$, we can obtain
\begin{equation}\label{eq:on_site_cocycle_covariance}
\nu_j({g},{h}) = \nu_{j-1}(\mathcal{T}({g}),\mathcal{T}({h}))
\end{equation}
up to coboundaries.

With $[\nu]\in H^2(G,U(1))$ nontrivial, we will need to revisit the earlier results. It is still true that the bond unitaries form projective representations of $G$, but the cocycles $\omega_j(g,h)$ are no longer translation invariant. A straightforward generalization of earlier arguments yields

\begin{equation}\label{eq:Grand_SPT_LSM}
{\omega_{j}(g,h)\omega_{j-1}^{-1}(g,h) } = \nu_{j}(g,h).
\end{equation}

For translational invariance, we also need 
\begin{equation}\label{eq:Grand_Pullback_Cocycle_Constraint_Generalized}
\begin{split}
v_j(g) &\doteq v_{j-1}(\mathcal{T}(g)) \\ 
\implies \omega_j(g,h) &= \omega_{j-1}(\mathcal{T}(g),\mathcal{T}(h)).
\end{split}
\end{equation}
These two equations are up to coboundaries. These constraints on bond cocycles in terms of the on-site cocycles agree with the results from Ref. \cite{pace_bulmash_2026_LSM_Modulated}. Using these constraints, one may systematically prove LSM theorems and 'SPT-LSM' theorems.
\subsection{Abelian exponential symmetries}
As an application, let us consider the general discrete Abelian group $\prod_I \mathbb{Z}_{N_I}$, where the generator $g_I$ of each $\Z_{N_I}$  is implemented by 
\begin{equation}
\mathcal{U}_I=\bigotimes^L_{j=1} U^{a^{j-1}_I}_{I}
\end{equation}
and $a_I$ is coprime to $N_I$. On each site, we further suppose the symmetry operators satisfies a possible projective representation:
\begin{equation}
U_I U_J=\exp(\frac{2\pi i n_{IJ}}{N_{IJ}}) U_J U_I
\end{equation}
with $I<J$, $N_{IJ}=\text{gcd} (N_I, N_J)$ and $n_{IJ}\in \Z_{N_{IJ}}$.

This gives rise to
\begin{equation}
    \nu_j(g_I,g_J)\nu^{-1}_j(g_J,g_I)=\exp(\frac{2\pi i n_{IJ} (a_I a_J)^{j-1}}{N_{IJ}}).
\end{equation}
Now suppose an injective MPS preserves this symmetry. The push-through condition then implies
\begin{equation}
  U^{a^{j-1}_I}_I \cdot A \doteq (v^{\dagger}_I)^{a^{j-1}_I} A v^{a^{j}_I}_I
\end{equation}

where $v_I$ acts on the virtual indices.

The projective representation on the virtual bond is  captured by the phase $\omega_j(g_I,g_J)\omega^{-1}_j(g_I,g_J)=v^{a^j_I}_I v^{a^j_J}_Jv^{-a^j_I}_I v^{-a^j_J}_J$ with $I<J$, which is invariant under coboundary transformations. Writing $\omega_0(g_I,g_J)\omega^{-1}_0(g_I,g_J)=v_I v_Jv^{-1}_I v^{-1}_J=\exp(2\pi i k_{IJ}/N_{IJ})$ where $k_{IJ}\in \Z_{N_{IJ}}$, this yields  $\omega_j(g_I,g_J)\omega^{-1}_j(g_I,g_J)=\exp( 2\pi i k_{IJ} (a_I a_J)^j/ N_{IJ})$.

Then according to Eq.~\eqref{eq:Grand_SPT_LSM}, we have
\begin{equation}
\begin{split}
    \omega_j(g_I,g_J)\omega^{-1}_j(g_I,g_J)=\omega_{j-1}(g_I,g_J)\omega^{-1}_{j-1}(g_I,g_J)\nu_j(g_I,g_J)\nu^{-1}_j(g_J,g_I),
    \end{split}
\end{equation}
which gives
\begin{equation}
     k_{IJ} [(a_I a_J)^{j}-(a_I a_J)^{j-1}]=n _{IJ}(a_I a_J)^{j-1} ~ (\text{mod} ~N_{IJ}).
\end{equation}
As both $a_I$ and $a_J$ are coprime to $N_{IJ}$ the above equation reduces to
\begin{equation}
    k_{IJ} (a_I a_J-1)=n_{IJ} ~ (\text{mod} ~N_{IJ}).
\end{equation}
The solutions are given by
\begin{equation}\label{eq: solution}
k_{IJ} =
\begin{cases}
\text{does not exist}& \text{if } c_{IJ} \text{ does not divides } n_{IJ}\\
\frac{n_{IJ}}{c_{IJ}} (\frac{a_I a_J-1}{c_{IJ}})^{-1}_{N_{IJ}/c_{IJ}}+\tau_{IJ} N_{IJ}/c_{IJ}& \text{if } c_{IJ} \text{ divides }n_{IJ},
\end{cases}
\end{equation}
where $c_{IJ}=\text{gcd}(a_I a_J-1,N_{IJ})$.
Here $(\cdots)^{-1}_{N_{IJ}/c_{IJ}}$ denotes the multiplicative inverse modulo $N_{IJ}/c_{IJ}$  and $\tau_{IJ}\in\Z_{c_{IJ}}$.  Therefore, there is an LSM constraint for this class of exponential symmetry when $c_{IJ}$ does not divides $n_{IJ}$.

Moreover, even when $c_{IJ}\mid n_{IJ}$, an obstruction to a trivial SPT phase may persist. If all allowed solutions $k_{IJ}$ are nonzero modulo $N_{IJ}$, the system necessarily realizes a nontrivial SPT phase. This leads to an SPT-LSM theorem—an obstruction to a trivial SPT. 
\subsection{Dipole symmetries}
Another application is the $\Z_N$ dipole symmetry generated by
\begin{equation}
    \mathcal{U}_q= \bigotimes^L_{j=1} U_{q,j},~ \mathcal{U}_d= \bigotimes^L_{j=1} U_{d,j},
\end{equation}
which satisfies:
\begin{equation}\label{eq:dipole aglebra}
    \mathcal{T}U_{d,j}\mathcal{T}^{\dagger}=U_{d,j-1}U_{q,j-1}.
\end{equation}
We further assume the possible projective representation on each site:
\begin{equation}\label{eq:dipole proj}
    U_{q,j}U_{d,j}=\exp(\frac{2\pi n i }{N})U_{d,j}U_{q,j}.
\end{equation}
Now suppose an injective MPS preserves this symmetry. The push-through condition then implies
\begin{equation}
  U_q \cdot A \doteq v^{\dagger}_q A v_q, \quad U_d \cdot A \doteq v^{\dagger}_{d,j-1} A v_{d,j}.
\end{equation}

Then the projective representation on the virtual bond is classified by the phase
\begin{equation}
    \omega_j(q,d)\omega^{-1}_j(d,q)=v_q v_{d,j}v^{-1}_q v^{-1}_{d,j} 
\end{equation}
which is invariant under coboundary transformations.
Furthermore due to Eq.~\eqref{eq:dipole aglebra}, we have
\begin{equation}
    v_{d,j-1}v_{q}\doteq v_{d,j}.
\end{equation}
 This yields
\begin{equation}
    \omega_j(q,d)\omega^{-1}_j(d,q)=\omega_{j-1}(q,d)\omega^{-1}_{j-1}(d,q),
\end{equation}
which implies that the projective representation on the virtual bond is site-independent.
When $n\ne 0$, this equation is incompatible with the projective representation on each site in Eq.~\eqref{eq:dipole proj}, which results in the LSM constraint. 

A concrete example for the dipole LSM is a $N$-dimensional qudit chain with symmetry operators
\begin{equation}
    \mathcal{U}_q=\bigotimes^L_{j=1} X_{j},  ~\mathcal{U}_d=\bigotimes^L_{j=1} Z^{-n}_jX^j_{j} .
\end{equation}
Here the $Z$ and $X$ are the on-site clock and shift operators. Under these symmetry actions, the qudit chain cannot be symmetric, gapped, and non-degenerate.

\section{Modulated Symmetry-Protected Exact Edge Degeneracies}\label{app: edge degeneracy}

An SPT phase realized on an open chain hosts topological edge modes which become degenerate as $L\to\infty$. At finite $L$, these edge degeneracies may be lifted by finite-size splittings of order $e^{-\gamma L}$.
However, it turns out that the modulation of the symmetry group leads to qualitatively distinct consequences for this finite-size behavior, in comparison with its unmodulated analogue. More precisely, the precise pattern of finite-size splittings of edge modes depends crucially on the modulation (or lack thereof) of the on-site symmetry group. 

To see this, consider an MPS tensor $A$ which obeys the generalized pushthrough rule for a modulated $G$-symmetry

\begin{equation}
\label{eq:general_push_through_app}
    U_g \cdot A \doteq \, v^\dagger(g)\, A\, v(\mathcal{T}(g)),
\end{equation}
and use it to build an MPS with arbitrary boundary conditions parametrized by $\mathcal{B}$:

\begin{equation}
\label{eq:general_bc_app}
\ket{\psi(\mathcal{B})} = \sum\Tr\biggr[\mathcal{B}A^{s_1}\cdots A^{s_L}\biggr]\ket{s_1\ldots s_L}.
\end{equation}
Here, periodic boundary conditions would have $\mathcal{B}=\mathbb{I}$, and open boundary conditions are parametrized by $\mathcal{B}=\bket{R}\bbra{L}$ where $R$ and $L$ label edge states. To generate a symmetric open MPS,  $\bbra{L},\bket{R}$ are required to transform in the projective representations of $G$ specified by $v^\dagger(g)$ and $v(\mathcal{T}^L(g))$.  The full state will be in a linear representation of $G$. \hfill \break

If $G$ is abelian, then for finite $L$, OBC states parametrized by Eq. \eqref{eq:general_bc_app} can be organized into one-dimensional irreducible representations. Thus there is no symmetry-based obstruction to completely lifting the edge degeneracy.   
By contrast, if $G$ is non-abelian, higher-dimensional linear irreducible representations may arise, allowing exactly degenerate edge-state multiplets to persist even at finite size.  
The key is that this organization of OBC states into linear $G$-representations depends crucially on the modulation automorphism $\mathcal{T}(\cdot)$. We illustrate this with an example. 

Consider $G = \mathbb{Z}_{2N}\rtimes \mathbb{Z}_2 = D_{4N}$ generated by $U_1$ and $U_2$ satisfying $U^{2N}_1=U^2_2=\mathbb{I}$ and $U_1 U_2=U_2 U^{-1}_1$. The SPT classification is $\mathbb{Z}_2$. One may verify that a valid bond projective representation of the nontrivial SPT phase is generated by a $\Z_{2n}$ operator $v_1= e^{i\theta\sigma_z/2}$ and a $\mathbb{Z}_2$ operator $v_2=\sigma_x$, where $\theta =\pi /N$. On a finite chain, we have edge states labelled by

$$\ket{\psi_{\alpha,\beta}} = \sum\Tr\biggr[\bket{\beta}\bbra{\alpha}A^{s_1}\cdots A^{s_L}\biggr]\ket{s_1\ldots s_L},$$ where $\alpha,\beta\in \pm 1$ label eigenvectors of $\sigma_z$. The symmetry charge under $U_1$ is $e^{i(\beta-\alpha)\theta/2}$ and applying $U_2$ flips this charge to $e^{-i(\beta-\alpha)\theta/2}$. The two states with $\beta=-\alpha$ thus transform in an irreducible 2-dimensional representation of $G$ and thus are degenerate for all $L$. We can take superpositions of the two states with $\alpha=\beta$ to form two 1-dimensional representations of $G$, so these two states generically get finite-size split into a basis of $U_2$ eigenstates.

Now suppose the $\mathbb{Z}_{2N}$ symmetry is exponentially modulated with an odd exponent base $b$, such that the same SPT can survive with the same OBC states $\ket{\psi_{\alpha\beta}}$. In this case, the bond projective representation acting on the right edge modes is generated by $(v_1)^{b^{L}}$  and $v_2$.
The symmetry charge of $\ket{\psi_{\alpha,\beta}}$ under $U_1$ is $e^{-i\alpha \theta/2}e^{ib^L\beta\theta/2}$, while the action of $U_2$ maps it to its partner  $\ket{\psi_{-\alpha,-\beta}}$ and flips this charge. As a result, unless $(b^L\beta-\alpha)$ is a multiple of $2N$,
the two states carry distinct $U_1$ charge and therefore form a two-dimensional irreducible representation of $G$, which guarantees their exact degeneracy. Accordingly, lifting this degeneracy requires $b^L+1$ or $b^L-1$ to be a multiple of $2N$. Finally, we note that if $\theta$ is promoted to a continuous parameter, corresponding to $G=U(1)\rtimes\Z_2$, then finite-size splittings would never emerge.

\bibliographystyle{unsrt} 
\bibliography{mspt}